\PassOptionsToPackage{table}{xcolor}
\documentclass[sigconf, nonacm]{acmart}

\setcopyright{none}
\copyrightyear{2027}
\acmYear{2027}
\acmDOI{XXXXXXX.XXXXXXX}
\acmISBN{978-1-4503-XXXX-X/2027/XX}
\acmConference[WWW '27]{Proceedings of the ACM Web Conference 2027}{May
  10--14, 2027}{Dublin, Ireland}
\acmBooktitle{Proceedings of the ACM Web Conference 2027 (WWW '27), May
  10--14, 2027, Dublin, Ireland}

\usepackage{tabularx}
\usepackage{multirow}
\usepackage{makecell}
\usepackage{enumitem}
\usepackage{algorithm}
\usepackage{algorithmic}
\usepackage{booktabs}
\usepackage{cuted}
\usepackage{marvosym} % \Letter -- corresponding-author marker
\usepackage[most]{tcolorbox}

\definecolor{promptlightgreen}{HTML}{EDF5EC}
\definecolor{promptdarkgreen}{HTML}{5B906B}
\definecolor{promptlightgrey}{HTML}{F2F2F2}
\definecolor{promptdarkgrey}{HTML}{666666}
\definecolor{stepred}{HTML}{B85450}
\definecolor{stepgreen}{HTML}{5B906B}
\newcommand{\stepone}{\textcolor{stepred}{\textbf{\textcircled{\scriptsize 1}}}}
\newcommand{\steptwo}{\textcolor{stepgreen}{\textbf{\textcircled{\scriptsize 2}}}}
\newcommand{\stepthree}{\textcolor{stepgreen}{\textbf{\textcircled{\scriptsize 3}}}}
\newcommand{\stepfour}{\textbf{\textcircled{\scriptsize 4}}}

\definecolor{panelBack}{HTML}{F1F4F7}
\definecolor{panelRule}{RGB}{130,136,145}
\newtcolorbox{topmatterBox}{
    enhanced,
    colback=panelBack,
    colframe=panelBack,
    boxrule=0pt,
    frame hidden,
    arc=3mm,
    boxsep=0pt,
    left=6mm, right=6mm, top=5mm, bottom=4mm,
    before skip=4mm, after skip=0mm,
    notitle
}
\newcommand{\headerlogo}{%
  \includegraphics[trim={32.8 43.5 23.4 42.3}, clip,
                   height=4mm]{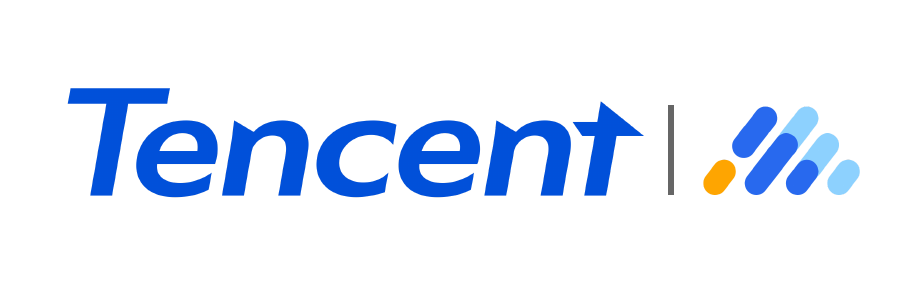}}
\newcommand{\panelnamefont}{\normalsize\bfseries}
\newcommand{\panelaffilfont}{\small}
\newcommand{\panelmailfont}{\ttfamily\footnotesize}
\makeatletter
\newbox\VerRec@titlebx
\let\@mktitle\@mktitle@i
\def\@titlefont{\Huge\bfseries}
\def\@mkauthors{\global\setbox\VerRec@titlebx=\box\mktitle@bx}
\def\@mkteasers{%
  \global\setbox\mktitle@bx=\vbox{%
    \hsize=\textwidth \linewidth=\textwidth \columnwidth=\textwidth
    \noindent\headerlogo\par
    \vskip 1.25mm
    {\color{panelRule}\hrule height 1.2pt}%
    \vskip 4.75mm
    \box\VerRec@titlebx
    \begin{topmatterBox}
      \setlength{\parindent}{0pt}%
      \authorpanel
      \medskip
      \noindent\ignorespaces\@abstract\par
      \ifx\@keywords\@empty\else
        \medskip\noindent{\bfseries\keywordsname:} \@keywords\par
      \fi
      \medskip
      {\color{panelRule}\hrule height 0.4pt}%
      \smallskip
      \noindent{\small\authornotepanel}\par
    \end{topmatterBox}
    \vskip 3.5mm % breathing room before the two-column body starts
  }%
  \global\let\VerRec@concepts\@concepts
  \global\let\VerRec@keywords\@keywords
  \global\let\@mkabstract\@empty
  \global\@ACM@printccsfalse
  \global\let\@concepts\@empty
  \global\let\@keywords\@empty}
\let\VerRec@printendtopmatter\@printendtopmatter
\def\@printendtopmatter{%
  \global\let\@concepts\VerRec@concepts
  \global\let\@keywords\VerRec@keywords
  \hypersetup{pdfsubject={\@concepts}, pdfkeywords={\@keywords}}%
  \VerRec@printendtopmatter}
\makeatother

\title[More Than a Score: Reinforcement Learning with Verbalized Feedback for Generative Recommendation]{More Than a Score: Reinforcement Learning with Verbalized Feedback for Generative Recommendation}

\newcommand{\authorpanel}{%
  {\panelnamefont
    \mbox{Zipeng Chen$^{1}$,}
    \mbox{Xinyu Lin$^{2}$\textsuperscript{\Letter},}
    \mbox{Yuliang Sun$^{3}$,}
    \mbox{Sheng Zhang$^{1}$,}
    \mbox{Zeyu Wang$^{1}$,}
    \mbox{Xiaowen Shi$^{1}$,}
    \mbox{Junwei Pan$^{1}$,}
    \mbox{Chengguo Yin$^{1}$}\par}
  \smallskip
  {\panelaffilfont
    \mbox{$^1$Tencent Inc., China}\quad
    \mbox{$^2$National University of Singapore, Singapore}\quad
    \mbox{$^3$The Chinese University of Hong Kong, Hong Kong, China}\par}
  {\panelmailfont
    $^1$beckzpchen@tencent.com, dragon\_201203@yeah.net, peterzywang@tencent.com,\\
    \hphantom{$^1$}princewen2019@outlook.com, jonaspan@tencent.com, turingyin@tencent.com\\
    $^2$xylin1028@gmail.com\\
    $^3$ylsun@link.cuhk.edu.hk\par}%
}
\newcommand{\authornotepanel}{%
  \textsuperscript{\Letter}\,Corresponding author.}

\renewcommand{\shortauthors}{Chen et al.}

\makeatletter
\gdef\authors{Zipeng Chen, Xinyu Lin, Yuliang Sun, Sheng Zhang, Zeyu Wang, Xiaowen Shi, Junwei Pan, Chengguo Yin}
\makeatother

\begin{abstract}
Large language models (LLMs) are increasingly used for generative recommendation, where recommended items are generated directly from users' interaction histories. To further improve recommendation quality, reinforcement learning (RL) has been applied to align these models with recommendation objectives. However, existing RL methods rely on scalar rewards to implicitly guide exploration. These rewards do not specify which aspects of a candidate conflict with the user's preferences or how the next candidate should differ. To address this gap, we propose VerRec, a verbalized feedback-guided RL framework for generative recommendation. Specifically, a user simulator generates personalized verbalized feedback based on the user's historical preferences and the candidate item profile. The policy uses this feedback to produce refined recommendations. Original and refined candidates form an expanded group and are jointly optimized with group relative policy optimization. The user simulator is only required during training and is removed at inference time. Experiments on three Amazon review datasets with two generative recommendation backbones demonstrate that VerRec improves recommendation accuracy and ranking quality. Further analyses show that verbalized guidance encourages exploration of more diverse candidates and helps the policy discover more rewarding recommendations during training.

\end{abstract}



\begin{document}

\maketitle

\section{Introduction}

\begin{figure}[t]
    \centering
    \captionsetup{skip=3pt}
    \includegraphics[width=\columnwidth]{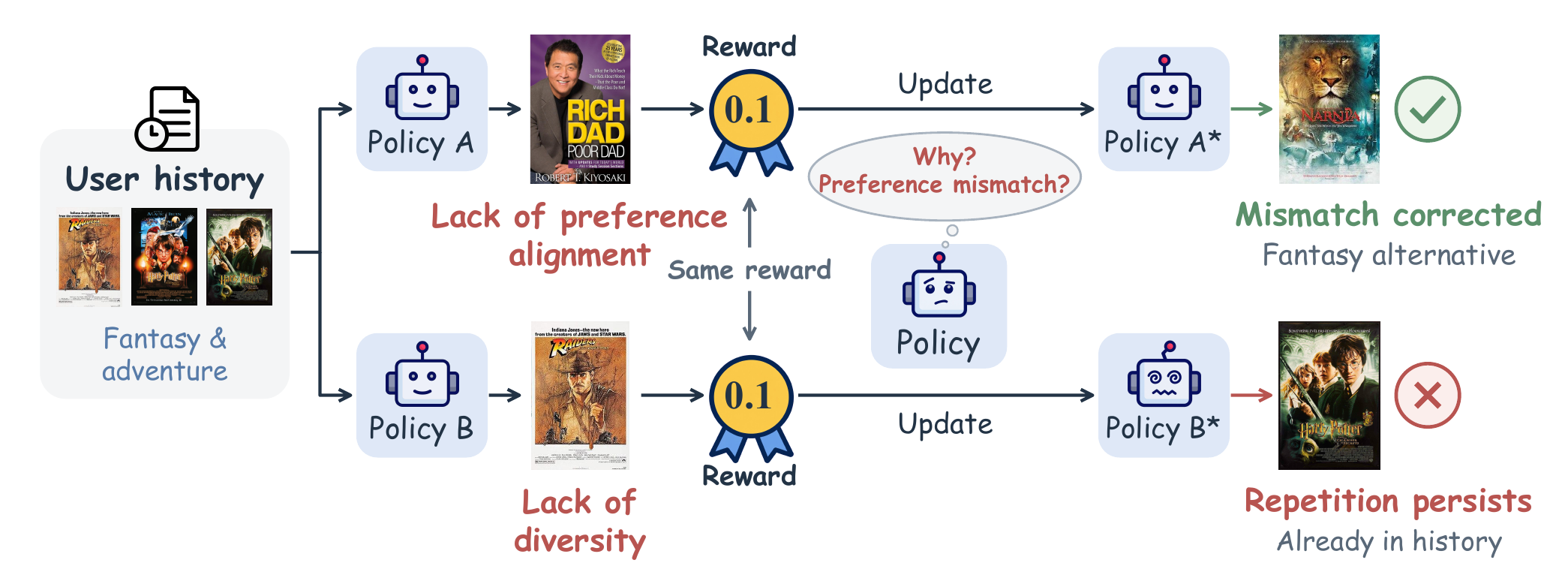}
    \caption{Ambiguity of scalar reward. Preference mismatch and repetitive recommendations receive the same low reward, although they require different corrections.}
    \label{fig:reward-ambiguity}
    \vspace{-6pt}
\end{figure}

\label{sec:intro}
% \xinyu{(1) The introduction of SFT and RL is not very natural, should highlight the existing standard paradigm used in both academia and industry, SFT -> RL paradigm; (2) at the end of this paragraph, should highlight the importance of RL in training generative recommender (better to list the benefits explicitly instead of just mentioning ``beyond SFT'').}

Generative recommendation (GR) reformulates personalized recommendation as autoregressive generation of the next item conditioned on a user's interaction history~\cite{rajput2023recommender,zheng2024adapting,wang2024learnable,feng2026fine}.
A common training pipeline combines supervised fine-tuning (SFT) with subsequent reinforcement learning (RL)~\cite{team2026onereason,zhou2025onerec,zhou2025openonerec,fan2026uncertainty}.
% \xinyu{The following two sentences can be combined into one.}
% Although SFT initializes the recommendation policy by learning to predict observed interactions, its supervision is largely restricted to observed behavior~\cite{ding2026towards,zhou2025onerec,wang2026gflowgr,zhou2025openonerec}; 
Although SFT trains the recommendation policy by fitting to the interactions, its supervision is largely restricted to observed behavior, causing poor generalization and diversity~\cite{ding2026towards,zhou2025onerec,wang2026gflowgr,zhou2025openonerec,lin2026bringing}. 
% RL subsequently complements SFT by optimizing the policy with rewards tied to recommendation objectives~\cite{kong2025minionerec,liu2025onerec,team2026onereason,zhou2025onerec,lin2026bringing}.
% RL complements this supervision by allowing the policy to explore candidate items and learn from the quality of its own recommendations.
% This makes RL an important stage for exploring candidates beyond observed interactions, aligning item generation with recommendation objectives, and improving recommendation accuracy and ranking quality~\cite{kong2025minionerec,tan2025rere,liu2025onerec,zhang2026reinforced,you2026r}.
RL then complements SFT by sampling items from the current policy and optimizing the policy with recommendation-oriented rewards~\cite{liu2025onerec,team2026onereason,zhou2025onerec}.
Exploring interactions that are absent from the training data, RL extends supervision beyond observed behavior and thereby improves the generalization ability, making it an important step in GR training~\cite{kong2025minionerec,tan2025rere,liu2025onerec,zhang2026reinforced,you2026r}.

RL for GR follows an iterative process. Conditioned on a user's interaction history, the policy samples candidate items and receives scalar rewards based on recommendation objectives. It then updates its parameters to increase the likelihood of candidates with higher rewards.
% RL for GR typically involves three steps:
% 1) \emph{Exploration:} The current policy samples candidate items conditioned on the user's interaction history.
% 2) \emph{Reward:} Each candidate receives a scalar reward based on recommendation objectives.
% 3) \emph{Policy update:} The policy is updated to increase the likelihood of candidates with higher rewards.
% \xinyu{I don't see how the following two types of work help understand the limitations of existing RL or the understanding of our method. I suggest discuss the existing types of work without indexing them. Otherwise, it seems like we are highlighting the two types of recommender algorithms, \ie with and without reasoning. By the way, do we have experiments on the reasoning-based methods that also equip our approach? if so, it is ok to mention two types of work here; otherwise, it is not that necessary.}
Prior work has applied this learning process to improve recommendation accuracy or user engagement through item generation~\cite{kong2025minionerec,tan2025rere,zhou2025onerec,zhou2025openonerec}, and to optimize reasoning trajectories together with item predictions based on recommendation outcomes~\cite{liu2025onerec,you2026r,team2026onereason}. In essence, these methods rely on scalar reward to judge the policy’s behaviors and update the policy to favor those that can lead to higher rewards.
% \xinyu{At the end of this paragraph, highlight the most crucial idea in RL is reward (for example, ``In essence, {RL uses rewards to judge the policy's behaviors and shifts the policy toward those with higher rewards}.'')}
% Although these designs provide increasingly refined supervision, they ultimately provide the model with only a scalar reward signal that conveys limited information.

% Despite encoding diverse recommendation objectives, these designs ultimately provide the recommender with only a scalar reward signal that conveys limited information.
% Such scalar reward signal makes on-policy exploration inherently inefficient: it tells the policy whether, or to what extent, a sampled recommendation is successful, but provides no \textit{explicit guidance} on \emph{why} the recommendation is unsuitable or \emph{how} it should be revised. The policy must therefore rely on costly trial-and-error sampling to discover rewarding recommendations by chance. 

% This reliance on scalar rewards leaves the direction of exploration \emph{implicit}: a low reward indicates an unsuccessful recommendation but leaves its cause of failure and the needed correction ambiguous.
However, the reliance on scalar rewards will lead to the \emph{\textbf{implicit exploration issue}}: a low reward can only indicate an unsuccessful recommendation but fail to reveal the underlying reason and optimization direction. 
% Figure~\ref{fig:reward-ambiguity} illustrates this ambiguity.
As an example shown in Figure~\ref{fig:reward-ambiguity}, 
we consider a user whose interaction history contains fantasy and adventure books.
Policy A recommends finance books that miss the user's interests, whereas Policy B recommends books the user has already read and offers insufficient diversity.
Both recommendations may receive the same low reward, but require different corrections: the former should move toward the user's preferred genres, while the latter should avoid previously consumed books within those genres.
The reward alone does not indicate which correction is needed, leaving the policy to discover better candidates through repeated trial and error.

Based on the above insights, we aim to investigate \emph{\textbf{explicit guidance}} in RL, which can identify the cause of failure and direct exploration toward suitable candidates. 
However, incorporating explicit guidance into RL for GR presents two key challenges:
% \xinyu{minor suggestion: consider two bullet points for guidance acquisition and guidance utilization, should be clearer for reader}: 
% 1) \emph{Guidance acquisition.} The guidance should identify why a candidate matches or conflicts with a user's preferences and indicate how it can be improved. Simulated verbalized feedback (VF) offers a potential source of such guidance by expressing preference assessments and refinement suggestions in natural language~\cite{cai2026mgfrec,zhang2025critique,huang2026bootstrapping}. The challenge is to obtain reliable, personalized feedback based on the user's historical preferences and the recommended candidate, so that it provides concrete directions for refinement.
% 2) \emph{Guidance utilization.} The obtained feedback should inform policy updates so that the policy can internalize the suggested improvements. However, verbalized feedback is discrete and cannot directly guide policy updates through backpropagation. The challenge is to incorporate the information in textual guidance into rewards, enabling the policy to learn from useful suggestions and improve subsequent recommendations.

\begin{itemize}[leftmargin=*,nosep]
    \item \textbf{Guidance acquisition.} The guidance should identify why a candidate matches or conflicts with a user's preferences and indicate how it can be improved. Simulated verbalized feedback (VF) offers a potential source of such guidance by expressing preference assessments and refinement suggestions in natural language~\cite{cai2026mgfrec,zhang2025critique,huang2026bootstrapping}. The challenge is to obtain reliable, personalized feedback based on the user's historical preferences and the recommended candidate, so that it provides concrete directions for refinement.
    \item \textbf{Guidance utilization.} The obtained feedback should inform policy updates so that the policy can internalize the suggested improvements. However, verbalized feedback is discrete and cannot directly guide policy updates through backpropagation. The challenge is to incorporate the information in textual guidance into rewards, enabling the policy to learn from useful suggestions and improve subsequent recommendations.
\end{itemize}

Building on these insights, we propose VerRec, a reinforcement learning framework with verbalized feedback for generative recommendation. At its core, VerRec consists of two tightly coupled components. 
1) \emph{Personalized Feedback Simulation}: For each original candidate generated by the policy, an LLM-based user simulator generates personalized VF based on the user's historical preferences and the candidate. This VF explains how the candidate matches or conflicts with the user's preferences.
2) \emph{Joint Optimization}: For each original candidate, the policy combines the candidate, its corresponding VF, and the user's history to construct a refinement prompt and generate a refined candidate. 
% \xinyu{minor suggestion: for the extended GRPO, can we give it a module/strategy name? for example, text-guided GRPO, this can also be used in the method section.}
The original and refined candidates are then combined into an expanded group and jointly optimized with GRPO using scalar rewards. This process allows VF to guide exploration while the reward function penalizes unsuccessful original candidates.

% 2) \emph{Feedback-Conditioned Refinement}: Each candidate and its corresponding VF are combined with the user histories to construct a refinement prompt. The policy then generates revised recommendations, producing rollouts guided by VF.
% 3) \emph{Joint Rollout Optimization}: For each selected candidate, the policy combines the candidate, its corresponding VF, and the user's history to construct a refinement prompt and generate a revised recommendation. Building on these insights, we propose VerRec, a VF-guided RL framework that explicitly transforms personalized VF into targeted exploration for GR. At its core, VerRec consists of two components.
To evaluate the effectiveness of VerRec, we conduct comprehensive experiments on three real-world datasets. The results demonstrate that VerRec effectively uses verbalized feedback to guide exploration and improve recommendation performance. Further analyses show that verbalized guidance encourages exploration of more diverse candidates and enables the policy to attain higher rewards as training progresses.
% helps the policy discover more rewarding recommendations during training\xinyu{What is rewarding recommendations...?}.
The main contributions of this work are summarized as follows:
\begin{itemize}[leftmargin=*,nosep]
    \item We formulate personalized verbalized feedback as an explicit exploration signal that guides the policy 
    % toward more diverse and relevant candidates 
    in reinforcement learning for generative recommendation.
    \item We propose VerRec, a novel RL framework that integrates personalized feedback simulation, feedback-conditioned candidate refinement, and joint GRPO optimization over original and refined candidates.
    \item We evaluate VerRec on three Amazon review datasets with two generative recommendation backbones. The results show improvements on recommendation performance and reveal how verbalized feedback  affects policy exploration and optimization.
\end{itemize}
\section{Related Work}
\label{sec:related}

\subsection{Generative Recommendation}
\label{sec:gr}

Generative recommendation has emerged as a paradigm for directly generating items from user interaction histories. 
Early work explores recommendation through text generation, using pretrained language models to process user and item information and generate recommendation outputs~\cite{geng2022recommendation,bao2025bi,zhang2026recommendation}. 
Subsequent work introduces compact semantic IDs (SIDs) to represent items and formulates next-item prediction as SID generation~\cite{kong2025minionerec,zhou2025onerec,wang2024learnable,rajput2023recommender}. 
Recent work further improves generative recommendation through multi-task supervised fine-tuning that aligns language semantics with collaborative signals~\cite{zheng2024adapting,he2026reasoning,ye2026align3gr}. 
Beyond supervised learning, recent studies apply reinforcement learning to align generative recommendation with recommendation objectives, either by directly optimizing item generation~\cite{kong2025minionerec,tan2025rere,zhang2026reinforced,ding2026towards} or by jointly optimizing reasoning and item prediction~\cite{liu2025onerec,you2026r,he2026reasoning,cai2026mgfrec}. 
These approaches use scalar rewards to evaluate recommendation outcomes and guide policy updates. Our work introduces personalized verbalized feedback to explicitly guide candidate exploration.

\subsection{Reinforcement Learning with Verbalized Feedback}
\label{sec:rlvf}

Verbalized (natural language) feedback provides actionable guidance beyond scalar rewards by identifying errors and suggesting improvements. 
Existing work has primarily explored such feedback in general-purpose tasks. One line of work uses such feedback at inference time, revising outputs through iterative self-critique or retaining verbal reflections to guide subsequent attempts without updating model parameters~\cite{madaan2023self,shinn2023reflexion}.
External tools can further ground this process by providing feedback for output verification and correction~\cite{gou2024critic,chen2024teaching}. Another line incorporates such feedback into model training, either by optimizing critique generators to improve a fixed model's revisions~\cite{akyurek2023rl4f,paul2024refiner} or by constructing synthetic preference data for offline optimization while constraining changes in unrelated contexts~\cite{stephan2024rlvf}. 
Recent studies combine textual critiques with scalar rewards in online RL, learning from initial responses and their refinements to guide exploration and using successful refinements as off-policy guidance~\cite{zhang2025critique,huang2026bootstrapping}. Our work investigates explicit guidance in RL for generative recommendation and develops a recommendation-specific framework for guidance acquisition and utilization.

\begin{figure*}[t]
    \centering
    \captionsetup{skip=3pt}
    \includegraphics[width=\textwidth]{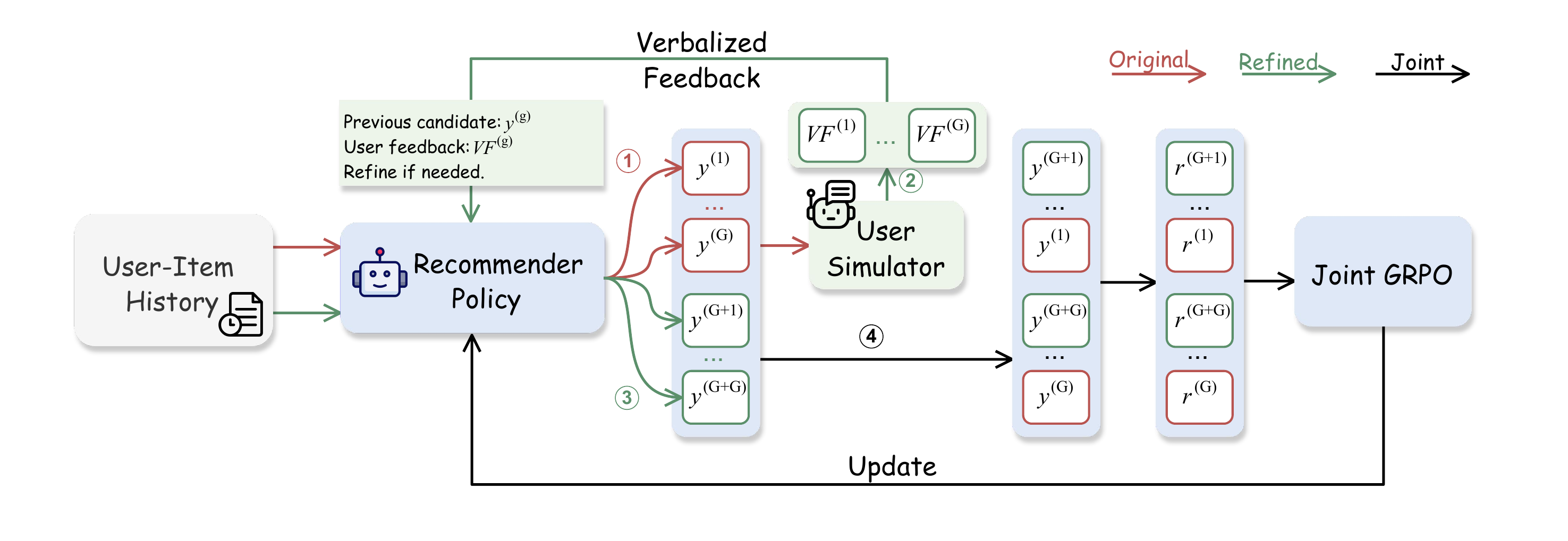}
    \caption{Overview of VerRec. \stepone~The recommender policy generates original SID candidates from the user--item interaction history. \steptwo~The user simulator analyzes the user's historical preferences and the candidate item profile to produce personalized verbalized feedback. \stepthree~Verbalized user feedback guides the refinement of original recommendation candidates. \stepfour~The original and refined candidates are combined into an expanded rollout group and jointly optimized with GRPO.}
    \label{fig:overview}
\end{figure*}

\subsection{User Simulator}
\label{sec:us}

User simulators provide controllable environments for training and evaluating recommender systems without costly online interactions. 
Traditional platforms model user choices or responses through handcrafted or learned behavioral models~\cite{ie2019recsim,zhao2023kuaisim,shi2019virtual}. 
Recent LLM-based simulators enrich this process with profiles, memory, and reasoning. 
One line of work directly initializes generative user agents with interaction histories and user profiles, relying on the in-context understanding ability of LLMs to simulate user preferences and behaviors~\cite{zhang2024generative,wang2025user,chen2025recusersim,zhang2026exploring,liu2026recoworld}. 
Another line fine-tunes LLM-based simulators on user feedback or behavioral trajectories, enabling them to better understand user preferences and align simulated behaviors with real users~\cite{chen2026beyond,zhang2026shop,wei2026mirroring}. 
Other studies evaluate whether LLMs can faithfully express preferences and provide consistent feedback in conversational recommendation~\cite{yoon2024evaluating,zhu2024reliable,chen2025recusersim}. 
Unlike prior simulators that mainly serve as interactive environments or evaluation tools, we employ a user simulator to synthesize personalized verbalized feedback from the user's historical preferences.

\section{Preliminary}
\label{sec:preliminary}

\subsection{Task Formulation}
\label{sec:task_formulation}

Generative recommendation aims to generate the next item that a user is likely to interact with based on their historical behaviors. Let $\mathcal{U}$ and $\mathcal{I}$ denote the sets of users and items, respectively. For each user $u\in\mathcal{U}$, the chronological interaction history is represented as $\mathcal{H}_u=(i_1^u,i_2^u,\ldots,i_T^u)$, where $i_T^u\in\mathcal{I}$. Following existing SID-based generative recommenders~\cite{rajput2023recommender,kong2025minionerec}, each item $i\in\mathcal{I}$ is represented as $y(i)=\operatorname{SID}(i)=(s_i^1,s_i^2,\ldots,s_i^L)$, where $s_i^l\in\mathcal{S}$ is the semantic token at level $l$, $\mathcal{S}$ is the SID token vocabulary, and $L$ is the SID length. The original LLM vocabulary is extended with the semantic tokens in $\mathcal{S}$, each treated as an indivisible token. Given $\mathcal{H}_u$, the generative recommender $\pi_\theta$, parameterized by $\theta$, is trained to directly generate the target SID $y_{T+1}^u$ for the next item $i_{T+1}^u\in\mathcal{I}$. We adopt the existing SID construction and SFT alignment pipeline~\cite{kong2025minionerec,wang2024learnable}, and focus on subsequent RL post-training.

 \subsection{Group Relative Policy Optimization}
\label{sec:grpo}

GRPO is an online RL algorithm that optimizes an LLM policy without training an additional value model~\cite{shao2024deepseekmath}. For each recommendation context $x_u$ constructed from $\mathcal{H}_u$, GRPO samples a group of $G$ candidate SIDs $\{y_u^{(g)}\}_{g=1}^{G}$ from the rollout policy $\pi_{\mathrm{old}}$ and assigns each candidate a scalar reward $r_u^{(g)}$. The group-relative advantage is computed by normalizing rewards within the same group:
\begin{equation}
    \hat{A}_u^{(g)}
    = \frac{r_u^{(g)}-\operatorname{mean}(\{r_u^{(j)}\}_{j=1}^{G})}
    {\operatorname{std}(\{r_u^{(j)}\}_{j=1}^{G})},
\end{equation}
where the same candidate-level advantage $\hat{A}_u^{(g)}$ is assigned to every token in $y_u^{(g)}$. GRPO maximizes the following token-level clipped surrogate objective:
\begin{equation}
    \begin{aligned}
        \mathcal{L}_{\mathrm{GRPO}}(\theta)
        ={}&\mathbb{E}\Bigg[\frac{1}{G}\sum_{g=1}^{G}\frac{1}{L}\sum_{t=1}^{L}
        \operatorname{CLIP}\big(\rho_{u,t}^{(g)}(\theta),\hat{A}_u^{(g)},\eta\big)\Bigg] \\
        &-\beta D_{\mathrm{KL}}\left(\pi_\theta\,\|\,\pi_{\mathrm{ref}}\right),
    \end{aligned}
\end{equation}
\begin{equation}
    \operatorname{CLIP}(\rho,A,\eta)
    = \min\big(\rho A,\,\operatorname{clip}(\rho,1-\eta,1+\eta)A\big),
\end{equation}
where the expectation is over recommendation contexts and candidate groups sampled from $\pi_{\mathrm{old}}$, and $L$ is the SID length. The parameters $\eta$ and $\beta$ control the clipping range and KL penalty, respectively, and $\rho_{u,t}^{(g)}(\theta)=\frac{\pi_\theta(y_{u,t}^{(g)}\mid x_u,y_{u,<t}^{(g)})}{\pi_{\mathrm{old}}(y_{u,t}^{(g)}\mid x_u,y_{u,<t}^{(g)})}$ is the token-level importance ratio, with $y_{u,<t}^{(g)}$ denoting the candidate prefix preceding token $t$. The KL term denotes the corresponding average token-level regularization against the reference policy $\pi_{\mathrm{ref}}$.
\section{Method}
\label{sec:method}

\subsection{Overview}
\label{sec:method_overview}

We propose VerRec, a reinforcement learning framework with verbalized feedback for generative recommendation. As illustrated in Figure~\ref{fig:overview}, VerRec consists of two components. In \textit{guidance acquisition} (Section~\ref{sec:user_feedback}), a user simulator provides personalized verbalized feedback for sampled candidates. In \textit{guidance utilization} (Section~\ref{sec:vf_rl}), the policy refines candidates using this feedback and jointly optimizes original and refined candidates with GRPO. The user simulator is only required during training.

\subsection{Guidance Acquisition}
\label{sec:user_feedback}

\begin{figure}[t]
    \centering
    \captionsetup{skip=3pt}
    \begin{tcolorbox}[
        width=\columnwidth,
        colback=promptlightgrey,
        colframe=promptdarkgrey,
        colbacktitle=promptdarkgrey,
        coltitle=white,
        title={User Simulator Prompt},
        fonttitle=\bfseries\large,
        arc=3mm,
        boxrule=1.2pt,
        left=5pt,
        right=5pt,
        top=5pt,
        bottom=5pt
    ]
        \footnotesize
        Here is information about yourself and a new \texttt{<item category>} ``\texttt{<item title>}'' being recommended to you. Ratings range from 1 to 5.\\[2pt]
        \textbf{\#\#\# Your Rating History \#\#\#}\\
        1. \texttt{<historical item title>}, \texttt{<rating>};\\
        \textbf{Positive Aspects:} [Aspect 1], [Aspect 2], \ldots\\
        \textbf{Negative Aspects:} [Aspect 1], [Aspect 2], \ldots\\
        \textbf{User Preference Elements:} [Preference 1], [Preference 2], \ldots\\
        \ldots\\
        \textbf{Your Average Rating:} \texttt{<user average rating>}\\[2pt]
        \textbf{\#\#\# Item Rating History by other users \#\#\#}\\
        1. \texttt{<rating>};\\
        \textbf{Positive Aspects:} [Aspect 1], [Aspect 2], \ldots\\
        \textbf{Negative Aspects:} [Aspect 1], [Aspect 2], \ldots\\
        \textbf{User Preference Elements:} [Preference 1], [Preference 2], \ldots\\
        \ldots\\
        \textbf{Item's Average Rating:} \texttt{<item average rating>}\\[2pt]
        \textbf{Task:} Based on the above information, provide your likely rating and review for ``\texttt{<item title>}''. The review should reflect your preferences and the item's characteristics.\\[2pt]
        \textbf{Output Format:}\\
        My Rating: [A rating between 1 and 5]\\
        My Review: [A review text]
    \end{tcolorbox}
    \caption{Prompt template for obtaining verbalized feedback from the user simulator.}
    \label{fig:user_simulator_prompt}
\end{figure}

Scalar rewards provide only a coarse correctness signal, without revealing how a recommended item aligns with the user's preferences. To provide more informative guidance, we introduce an LLM-based user simulator that generates personalized verbalized feedback for sampled candidates. Such feedback explicitly describes the user's interests and helps the recommendation policy explore better candidates.

Specifically, to obtain a fine-grained representation of user preferences, following prior work~\cite{fang2026reason4rec}, we use an LLM to summarize the reasons behind each historical interaction into positive and negative aspects. We further characterize the user's rating tendency by computing the average rating over all previous interactions. To provide a detailed profile of the candidate item, we collect its reviews from other users and summarize them in the same manner, together with its historical average rating. For both sides, we retain the ten most recent records before the current timestamp to prevent information leakage. Formally, given the resulting user history $\mathcal{H}_u$, candidate item $i_T$, and item review history $\mathcal{E}_T$, the user simulator generates simulated verbalized feedback $VF_T$:
\begin{equation}
    f_U(\mathcal{H}_u,i_T,\mathcal{E}_T)=VF_T.
    \label{eq:vf}
\end{equation}
$VF_T$ consists of a rating that directly reflects the user’s degree of preference for the candidate item and a review that explains the specific reasons behind this preference. The specific user simulator prompt is presented in Figure~\ref{fig:user_simulator_prompt}.

\subsection{Guidance Utilization}
\label{sec:vf_rl}

We now describe how verbalized feedback is used to optimize the recommendation policy $\pi_\theta$. The goal is to let the policy internalize feedback-driven improvements during training, so that it can generate better recommendations at inference time without requiring the user simulator.

\subsubsection{Feedback-Conditioned Refinement}
\label{sec:refinement}

Given a recommendation context $x_u$ constructed from $\mathcal{H}_u$ (as shown in Figure~\ref{fig:original_prompt}), the policy first samples a group of $G$ original candidates $\{y_{u,0}^{(g)}\}_{g=1}^{G}$:
\begin{equation}
    y_{u,0}^{(g)}\sim\pi_{\mathrm{old}}(\cdot\mid x_u),\qquad g=1,\ldots,G.
\end{equation}
For each original candidate $y_{u,0}^{(g)}$, the corresponding item is passed to the user simulator, which returns personalized verbalized feedback $VF_u^{(g)}$ following Eq.~\eqref{eq:vf}. Conditioned on the same history, the original recommendation, and this feedback, the policy generates a refined candidate:
\begin{equation}
    y_{u,1}^{(g)}\sim\pi_{\mathrm{old}}(\cdot\mid x_u,y_{u,0}^{(g)},VF_u^{(g)}).
\end{equation}
Both $y_{u,0}$ and $y_{u,1}$ are produced by the same model; they differ only in the generation context. The original prompt asks the policy to predict the next item from the user's interaction history, while the refined prompt further includes the previous recommendation and the simulated user feedback, as shown in Figures~\ref{fig:original_prompt} and~\ref{fig:refined_prompt}. Constrained decoding is applied in both cases, so that every completed SID corresponds to a valid item.

\begin{figure}[t]
    \centering
    \captionsetup{skip=3pt}
    \begin{tcolorbox}[
        width=\columnwidth,
        colback=promptlightgrey,
        colframe=promptdarkgrey,
        colbacktitle=promptdarkgrey,
        coltitle=white,
        title={Original Recommendation Prompt},
        fonttitle=\bfseries\large,
        arc=3mm,
        boxrule=1.2pt,
        left=5pt,
        right=5pt,
        top=5pt,
        bottom=5pt
    ]
        \footnotesize
        \textbf{\#\#\# User Input:}\\
        The user has interacted with items \texttt{<SID$_1$>, <SID$_2$>, \ldots} in chronological order. Can you predict the next possible item that the user may expect?\\[2pt]
        \textbf{\#\#\# Response:}\\
        \texttt{<semantic ID>}
    \end{tcolorbox}
    \caption{Prompt template for generating original recommendation candidates.}
    \label{fig:original_prompt}
\end{figure}

\begin{figure}[t]
    \centering
    \captionsetup{skip=3pt}
    \begin{tcolorbox}[
        width=\columnwidth,
        colback=promptlightgrey,
        colframe=promptdarkgrey,
        colbacktitle=promptdarkgrey,
        coltitle=white,
        title={Refinement Prompt},
        fonttitle=\bfseries\large,
        arc=3mm,
        boxrule=1.2pt,
        left=5pt,
        right=5pt,
        top=5pt,
        bottom=5pt
    ]
        \footnotesize
        Your previous task and recommendation was:...\\
        \texttt{<original recommendation prompt>}\\
        \texttt{<original semantic ID>}\\[2pt]
        User's feedback on your recommendation: \texttt{<verbalized feedback>}\\[2pt]
        Based on the above feedback, please refine your recommendation (if needed). Can you recommend a suitable next item for the user?\\[2pt]
        \textbf{\#\#\# Response:}\\
        \texttt{<refined semantic ID>}
    \end{tcolorbox}
    \caption{Prompt template for generating feedback-conditioned refined candidates.}
    \label{fig:refined_prompt}
\end{figure}

% \subsubsection{Joint Rollout Optimization}
% \label{sec:joint_opt}

% For each context $x_u$, we sample a group of $G$ original candidates $\{y_{g,0}\}_{g=1}^{G}$. For each original candidate, we obtain verbalized feedback and generate a corresponding refined candidate $\{y_{g,1}\}_{g=1}^{G}$. The original and refined recommendations are then assembled into an expanded group of size $2G$ and ranked by their generation scores $s(y)=\sum_{t=1}^{|y|}\log \pi_{old}\!\left(y_t\mid x_u,y_{<t}\right)$:
% \begin{equation}
%     \mathcal{G}(x_u)=\operatorname{Rank}\big(\{y_{g,0}\}_{g=1}^{G}\cup\{y_{g,1}\}_{g=1}^{G};\,s(\cdot)\big).
% \end{equation}
% We then apply the clipped GRPO objective from Section~\ref{sec:grpo} to the expanded group:
% \begin{equation}
%     \begin{aligned}
%     \mathcal{L}_{\textsc{TellMeWhy}}(\theta)
%     ={}&\mathbb{E}_{x_u}\left[
%     \frac{1}{2G}\sum_{y\in\mathcal{G}(x_u)}\frac{1}{L}\sum_{t=1}^{L}
%     \operatorname{CLIP}\big(\rho_{\theta,t}(y),A(y),\eta\big)
%     \right]\\
%     &-\beta D_{\mathrm{KL}}\left(\pi_\theta\,\|\,\pi_{\mathrm{ref}}\right),
%     \end{aligned}
% \end{equation}
% Here, $\rho_{\theta,t}(y)=\frac{\pi_\theta(y_t\mid x_u,y_{<t})}{\pi_{\mathrm{old}}(y_t\mid x_u,y_{<t})}$ is the token-level policy ratio. In this way, a rewarding refined candidate can break an otherwise uninformative group and provide a useful learning signal for both successful revisions and failed original rollouts. After training, the user simulator and the refinement prompt are discarded, and the policy generates recommendations using only the original prompt.

\subsubsection{Reward Design}
\label{sec:reward}

After refinement, we combine the original and refined candidates and rescore each candidate using $s(y)=\sum_{t=1}^{L}\log\pi_{\mathrm{old}}\!\left(y_t\mid x_u,y_{<t}\right)$. We then rerank the candidates in descending order of $s(y)$ to form the final expanded rollout group $\mathcal{G}(x_u)$. To align the policy with recommendation objectives, we assign each candidate $y\in\mathcal{G}(x_u)$ a combination of a rule-based reward and a rank-aware reward. Let $y_{T+1}^{u}$ denote the target SIDs. The rule-based term provides a binary correctness signal:
\begin{equation}
    r_{\mathrm{rule}}(y)=\mathbb{I}\!\left[y=y_{T+1}^{u}\right].
\end{equation}
A purely binary reward treats all incorrect candidates equally. We therefore further penalize hard negatives according to their ranks $\rho(y)$ within the expanded group, where $\rho(y)=1$ corresponds to the most probable candidate:
\begin{equation}
    \widetilde{r}_{\mathrm{rank}}(y)=
    \begin{cases}
        0, & \text{if } y=y_{T+1}^{u},\\
        -\dfrac{1}{\log(\rho(y)+1)}, & \text{otherwise.}
    \end{cases}
\end{equation}
The rank-aware reward is then normalized over the $2G$ candidates,
\begin{equation}
    r_{\mathrm{rank}}(y)= -\frac{\widetilde{r}_{\mathrm{rank}}(y)}{\sum_{y'\in\mathcal{G}(x_u)}\widetilde{r}_{\mathrm{rank}}(y')},
\end{equation}
and the final reward is
\begin{equation}
    r(y)=r_{\mathrm{rule}}(y)+r_{\mathrm{rank}}(y).
\end{equation}

\subsubsection{Joint Optimization}
\label{sec:joint_opt}

We jointly optimize the original and refined candidates to learn from both direct exploration and feedback-conditioned refinement. For an expanded rollout group containing $G$ original and $G$ refined candidates, the final objective is
\begin{equation}
    \begin{aligned}
    \mathcal{L}_{VerRec}(\theta)
    ={}&\mathbb{E}\Bigg[
    \frac{1}{2G}\sum_{g=1}^{G}
    \bigl(\mathcal{J}_{u,0}^{(g)}+\mathcal{J}_{u,1}^{(g)}\bigr)
    \Bigg]\\
    &-\beta D_{\mathrm{KL}}\left(\pi_\theta\,\|\,\pi_{\mathrm{ref}}\right),
    \end{aligned}
    \label{eq:joint_objective}
\end{equation}
where $\mathcal{J}_{u,0}^{(g)}$ and $\mathcal{J}_{u,1}^{(g)}$ use different surrogates for original and refined candidates. We apply the CLIP operator only to the original candidate samples, not to the candidates refined by feedback:
\begin{equation}
    \mathcal{J}_{u,0}^{(g)}
    =\frac{1}{L}\sum_{t=1}^{L}
    \operatorname{CLIP}\bigl(
    \rho_{u,0,t}^{(g)}(\theta),\hat{A}_{u,0}^{(g)},\eta
    \bigr),
    \label{eq:joint_surrogate}
\end{equation}
\begin{equation}
    \mathcal{J}_{u,1}^{(g)}
    =\frac{1}{L}\sum_{t=1}^{L}
    f\bigl(p_{u,1,t}^{(g)}(\theta)\bigr)\hat{A}_{u,1}^{(g)}.
    \label{eq:joint_refined_surrogate}
\end{equation}
To compute the advantages, we first calculate the reward of each candidate in the final expanded candidate group $\mathcal{G}(x_u)$ and collect these rewards as $\mathbf{r}(x_u)$. We then normalize these rewards as follows:
\begin{equation}
    \hat{A}_{u,b}^{(g)}
    =\frac{r(y_{u,b}^{(g)})-\operatorname{mean}(\mathbf{r}(x_u))}{\operatorname{std}(\mathbf{r}(x_u))},
    \qquad b\in\{0,1\},
    \label{eq:joint_advantage}
\end{equation}
Since refined candidates are generated with feedback but optimized under the original context $x_u$, they may have very low token probabilities under the optimization policy. 
This mismatch can lead to unstable probability ratios and high-variance updates, potentially destabilizing training.
To address this issue, inspired by LUFFY~\cite{yan2026learning}, we employ Policy Shaping $f(x)=\frac{x}{x+\gamma}$, where $0<\gamma<1$.
This shaping function gives low-probability tokens in refined samples greater relative emphasis, encouraging the policy to learn from rewarding refinements that are rarely generated under the original context. The standard ratio for original candidates and the shaped weight for refined candidates are defined as follows:
\begin{equation}
    \rho_{u,0,t}^{(g)}(\theta)
    =\frac{\pi_\theta(y_{u,0,t}^{(g)}\mid x_u,y_{u,0,<t}^{(g)})}
    {\pi_{\mathrm{old}}(y_{u,0,t}^{(g)}\mid x_u,y_{u,0,<t}^{(g)})},
    \label{eq:joint_ratio}
\end{equation}
\begin{equation}
    f\bigl(p_{u,1,t}^{(g)}(\theta)\bigr)
    =\frac{\pi_\theta(y_{u,1,t}^{(g)}\mid x_u,y_{u,1,<t}^{(g)})}
    {\pi_\theta(y_{u,1,t}^{(g)}\mid x_u,y_{u,1,<t}^{(g)})+\gamma},
    \label{eq:joint_shaping}
\end{equation}
\section{Experiments}
\label{sec:exp}

In this section, we present the experimental results and corresponding analysis to answer the following research questions (RQs).
\begin{itemize}[leftmargin=*,nosep]
    \item \textbf{RQ1:} How does the recommendation performance of VerRec compare to other baselines?
    \item \textbf{RQ2:} How do different feedback strategies compare in simulation quality and downstream recommendation performance?
    \item \textbf{RQ3:} How does verbalized feedback affect candidate exploration during RL training?
    \item \textbf{RQ4:} How do the key components of VerRec affect its overall performance?
\end{itemize}

\subsection{Experimental Setup}
\label{sec:exp_settings}

\subsubsection{Datasets and Evaluation}
\label{sec:datasets}

\begin{table}[t]
  \centering
  \captionsetup{skip=3pt}
  \caption{Statistics of the recommendation datasets.}
  \label{tab:dataset_stats}
  \small
  \begin{tabular}{lrrr}
    \toprule
    \textbf{Dataset} & \textbf{\#Users} & \textbf{\#Items} & \textbf{\#Interaction} \\
    \midrule
    Book & 12,969 & 13,515 & 69,794 \\
    Baby & 17,734 & 6,863 & 128,755 \\
    Cell Phones & 26,482 & 9,616 & 151,087 \\
    \bottomrule
  \end{tabular}
\end{table}

Following prior works~\cite{bao2024decoding,bao2025bi,kong2025minionerec}, we conduct experiments on three categories of the Amazon review dataset\footnote{
\url{https://cseweb.ucsd.edu/~jmcauley/datasets/amazon/links.html}.}: Cell Phones and Accessories (Cell Phones), Baby, and Book. Each dataset contains users' interaction histories and rich textual metadata associated with the items. We apply 5-core filtering to ensure that every user and item has at least five interactions. Each user's interactions are chronologically ordered and segmented using a sliding window with a maximum sequence length of 10. To simulate realistic recommendation scenarios, we adopt a temporal split and partition the resulting sequences into training, validation, and test sets at a ratio of 8:1:1, reserving the most recent interactions for validation and testing. Detailed statistics for each category are reported in Table~\ref{tab:dataset_stats}.

For evaluation metrics, we adopt the full-item ranking setting, where each model is required to rank all candidate items rather than a sampled subset. We evaluate the effectiveness of top-$K$ recommendations using Hit Ratio (H@$K$) and Normalized Discounted Cumulative Gain (N@$K$), with $K\in\{5,10,20\}$.

\begin{table*}[t]
  \centering
  \captionsetup{skip=3pt}
  \caption{Overall performance of VerRec and baselines on three real-world datasets. All results are averaged over 5 runs with different random seeds. The best results are in bold, and the second-best results (i.e., the strongest baseline) are underlined. Gray rows denote the SFT backbone. $^{*}$ indicates that the improvement of VerRec over the strongest baseline is statistically significant (paired $t$-test, $p<0.05$).}
  \label{tab:recommendation_results}
  \small
  \setlength{\tabcolsep}{1pt}
  \setlength{\aboverulesep}{0.35pt}
  \setlength{\belowrulesep}{0.35pt}
  \renewcommand{\arraystretch}{1.15}
  \resizebox{\textwidth}{!}{%
  \begin{tabular}{l|cccccc|cccccc|cccccc}
    \toprule
    \multirow{2}{*}{Model}
      & \multicolumn{6}{c|}{Cell Phones}
      & \multicolumn{6}{c|}{Baby}
      & \multicolumn{6}{c}{Book} \\
    \cmidrule(lr){2-7}\cmidrule(lr){8-13}\cmidrule(lr){14-19}
      & H@5 & N@5 & H@10 & N@10 & H@20 & N@20
      & H@5 & N@5 & H@10 & N@10 & H@20 & N@20
      & H@5 & N@5 & H@10 & N@10 & H@20 & N@20 \\
    \midrule
    SASRec  & 0.0189 & 0.0114 & 0.0295 & 0.0148 & 0.0456 & 0.0188 & 0.0100 & 0.0065 & 0.0199 & 0.0097 & 0.0336 & 0.0131 & 0.0348 & 0.0233 & 0.0527 & 0.0291 & 0.0762 & 0.0356 \\
    GRU4Rec & 0.0238 & 0.0160 & 0.0394 & 0.0209 & 0.0611 & 0.0264 & 0.0126 & 0.0077 & 0.0244 & 0.0115 & 0.0359 & 0.0147 & 0.0209 & 0.0142 & 0.0337 & 0.0183 & 0.0576 & 0.0243 \\
    Caser   & 0.0262 & 0.0162 & 0.0420 & 0.0213 & 0.0639 & 0.0274 & 0.0103 & 0.0063 & 0.0203 & 0.0095 & 0.0349 & 0.0135 & 0.0299 & 0.0184 & 0.0504 & 0.0250 & 0.0752 & 0.0319 \\
    \midrule
    BIGRec  & 0.0290 & 0.0205 & 0.0415 & 0.0244 & 0.0578 & 0.0286 & 0.0125 & 0.0081 & 0.0215 & 0.0109 & 0.0333 & 0.0139 & 0.0347 & 0.0237 & 0.0512 & 0.0289 & 0.0696 & 0.0340 \\
    D$^3$   & 0.0330 & 0.0226 & 0.0464 & 0.0268 & 0.0646 & 0.0314 & 0.0142 & 0.0089 & 0.0236 & 0.0119 & 0.0373 & 0.0152 & 0.0395 & 0.0262 & 0.0572 & 0.0318 & 0.0777 & 0.0375 \\
    S-DPO   & 0.0317 & 0.0213 & 0.0457 & 0.0258 & 0.0636 & 0.0307 & 0.0138 & 0.0092 & 0.0232 & 0.0122 & 0.0376 & 0.0155 & 0.0383 & 0.0261 & 0.0562 & 0.0324 & 0.0772 & 0.0382 \\
    MiniOneRec  & \underline{0.0346} & \underline{0.0240} & 0.0477 & 0.0282 & 0.0664 & \underline{0.0330} & \underline{0.0149} & \underline{0.0095} & \underline{0.0247} & \underline{0.0126} & 0.0383 & \underline{0.0160} & 0.0416 & 0.0278 & 0.0587 & 0.0334 & 0.0800 & 0.0393 \\
    \midrule
    LatentR$^3$ & 0.0338 & 0.0231 & 0.0491 & 0.0280 & 0.0655 & 0.0323 & 0.0145 & 0.0093 & 0.0242 & 0.0124 & 0.0384 & 0.0157 & 0.0417 & 0.0279 & 0.0595 & 0.0335 & 0.0797 & 0.0391 \\
    R$^2$ec     & \underline{0.0346} & 0.0233 & \underline{0.0503} & \underline{0.0284} & \underline{0.0671} & 0.0326 & 0.0146 & 0.0092 & 0.0240 & 0.0122 & \underline{0.0388} & 0.0159 & \textbf{0.0423} & \underline{0.0283} & \underline{0.0600} & \underline{0.0339} & \underline{0.0802} & \underline{0.0395} \\
    \midrule
    \rowcolor{gray!15}
    MiniOneRec-SFT
      & 0.0273 & 0.0166 & 0.0462 & 0.0227 & 0.0644 & 0.0277 & 0.0137 & 0.0090 & 0.0234 & 0.0121 & 0.0359 & 0.0153 & 0.0385 & 0.0257 & 0.0567 & 0.0321 & 0.0795 & 0.0384 \\
    \quad \textbf{VerRec}
      & \textbf{0.0357}$^{*}$ & \textbf{0.0246}$^{*}$ & \textbf{0.0525}$^{*}$ & \textbf{0.0300}$^{*}$ & \textbf{0.0730}$^{*}$ & \textbf{0.0351}$^{*}$ & \textbf{0.0159}$^{*}$ & \textbf{0.0099}$^{*}$ & \textbf{0.0279}$^{*}$ & \textbf{0.0137}$^{*}$ & \textbf{0.0442}$^{*}$ & \textbf{0.0178}$^{*}$ & 0.0416 & \textbf{0.0290}$^{*}$ & \textbf{0.0623}$^{*}$ & \textbf{0.0357}$^{*}$ & \textbf{0.0820}$^{*}$ & \textbf{0.0406}$^{*}$ \\
    \midrule
    \rowcolor{gray!15}
    LETTER
      & 0.0260 & 0.0188 & 0.0420 & 0.0239 & 0.0646 & 0.0295 & 0.0135 & 0.0084 & 0.0222 & 0.0112 & 0.0359 & 0.0150 & 0.0374 & 0.0262 & 0.0526 & 0.0316 & 0.0797 & 0.0391 \\
    \quad \textbf{VerRec}
      & \textbf{0.0361}$^{*}$ & \textbf{0.0264}$^{*}$ & \textbf{0.0513}$^{*}$ & \textbf{0.0313}$^{*}$ & \textbf{0.0733}$^{*}$ & \textbf{0.0368}$^{*}$ & \textbf{0.0163}$^{*}$ & \textbf{0.0107}$^{*}$ & \textbf{0.0253}$^{*}$ & \textbf{0.0136}$^{*}$ & \textbf{0.0397}$^{*}$ & \textbf{0.0169}$^{*}$ & 0.0420 & \textbf{0.0311}$^{*}$ & \textbf{0.0609}$^{*}$ & \textbf{0.0372}$^{*}$ & \textbf{0.0822}$^{*}$ & \textbf{0.0425}$^{*}$ \\
    \bottomrule
  \end{tabular}%
  }
\end{table*}

\subsubsection{Baselines}
\label{sec:baselines}

We compare VerRec with three groups of methods: traditional recommenders, LLM-based recommenders without reasoning, and reasoning-enhanced LLM-based recommenders.

The traditional recommenders are as follows:
\begin{itemize}[leftmargin=*,itemsep=0pt,topsep=4pt,parsep=0pt,partopsep=0pt,beginpenalty=10000]
    \item \textit{SASRec}~\cite{kang2018self} uses self-attention to weight historical interactions according to their relevance to the next item.
    \item \textit{GRU4Rec}~\cite{hidasi2015session} encodes interaction sequences with gated recurrent units and predicts the next item from the evolving hidden state.
    \item \textit{Caser}~\cite{tang2018personalized} applies horizontal and vertical convolutional filters to a matrix of recent item embeddings to extract sequential patterns.
\end{itemize}

The LLM-based recommenders without reasoning are as follows:
\begin{itemize}[leftmargin=*,itemsep=0pt,topsep=4pt,parsep=0pt,partopsep=0pt,beginpenalty=10000]
    \item \textit{BIGRec}~\cite{bao2025bi} fine-tunes an LLM to generate textual item representations from interaction histories, then maps the generated outputs to items in the catalog.
    \item \textit{D$^3$}~\cite{bao2024decoding} adjusts decoding to reduce score amplification and repetitive recommendations. It removes length normalization and incorporates guidance from a text-free recommender.
    \item \textit{S-DPO}~\cite{chen2024softmax} extends preference optimization to contrast a preferred item with multiple negative items through a softmax objective, strengthening ranking supervision.
    \item \textit{LETTER}~\cite{wang2024learnable} learns item codes that combine hierarchical semantics, collaborative information, and diverse code assignments for generative recommendation.
    \item \textit{MiniOneRec}~\cite{kong2025minionerec} generates semantic IDs and follows supervised fine-tuning with GRPO. It combines constrained beam search with accuracy and rank-aware rewards.
\end{itemize}

The reasoning-enhanced LLM-based recommenders are as follows:
\begin{itemize}[leftmargin=*,itemsep=0pt,topsep=4pt,parsep=0pt,partopsep=0pt,beginpenalty=10000]
    \item \textit{R$^2$ec}~\cite{you2026r} combines explicit reasoning with a dedicated item prediction head. Its RL framework jointly optimizes reasoning and recommendation using ranking and similarity rewards.
    \item \textit{LatentR$^3$}~\cite{zhang2026reinforced} represents reasoning with continuous latent tokens and optimizes them through RL. It uses negative target-item perplexity as the reward without generating reasoning traces.
\end{itemize}

We evaluate VerRec on two backbones, \textbf{MiniOneRec-SFT} and \textbf{LETTER}, and report their performance before applying our framework. MiniOneRec-SFT is the supervised checkpoint before MiniOneRec's RL stage~\cite{kong2025minionerec}.

\subsubsection{Implementation Details}
\label{sec:details}
For MiniOneRec-SFT and LETTER, we construct item SIDs following their respective tokenization schemes~\cite{kong2025minionerec,wang2024learnable}. For both backbones, we perform SFT on Qwen2.5-3B-Instruct~\cite{qwen2.5}.  The learning rates are $3\times10^{-4}$ for SFT and $1\times10^{-5}$ for RL. Both stages use AdamW with a batch size of 128. For the key RL hyperparameters, we set $\gamma = 0.1$ and $\beta = 0.001$.
We adopt Qwen3-30B-A3B-Instruct-2507~\cite{yang2025qwen3} both to summarize user and item preferences from historical reviews and to serve as the user simulator that generates verbalized feedback.
In the performance comparison with existing recommenders, traditional recommenders are trained with binary cross-entropy and the Adam optimizer; learning rates are chosen from $\{10^{-2}, 10^{-3}, 10^{-4}\}$, and the weight-decay term is scanned over $\{10^{-2}, 10^{-3}, 10^{-4}, 10^{-5}, 10^{-6}\}$.
For LLM-based baselines, we carefully tune the hyperparameters of the baselines following the suggestions in the original papers.

\subsection{Overall Performance (RQ1)}
\label{sec:overall_performance}

Table~\ref{tab:recommendation_results} reports the overall recommendation performance under the full-item ranking setting. VerRec outperforms all other baselines on most metrics across the evaluated datasets with both recommendation backbones. Its advantages extend to comparisons with both direct-generation and reasoning-enhanced recommenders, demonstrating the effectiveness of the proposed framework in improving recommendation accuracy and ranking quality.

Compared with MiniOneRec, VerRec achieves further improvements on most metrics from the same MiniOneRec-SFT checkpoint. This comparison supports the effectiveness of our  novel training framework beyond existing RL methods that rely solely on scalar rewards. Personalized verbalized feedback provides explicit guidance for refinement, while scalar rewards evaluate the original and refined candidates for joint policy optimization.

\subsection{Feedback Quality and Effect (RQ2)}
\label{sec:feedback_quality}

\begin{table}[t]
  \centering
  \captionsetup{skip=3pt}
  \caption{Performance comparison on the three Amazon review datasets. Lower MAE and RMSE indicate better performance, while higher BLEURT indicates better performance.}
  \label{tab:main_results}
  \small
  \renewcommand{\arraystretch}{0.95}
  \setlength{\aboverulesep}{0.35ex}
  \setlength{\belowrulesep}{0.35ex}
  \setlength{\tabcolsep}{0pt}
  \begin{tabular*}{0.95\columnwidth}{@{\extracolsep{\fill}}llccc@{}}
    \toprule
    \textbf{Dataset} & \textbf{Method}
      & \textbf{MAE}$\downarrow$
      & \textbf{RMSE}$\downarrow$
      & \textbf{BLEURT}$\uparrow$ \\
    \midrule
    \multirow{3}{*}{Cell Phones}
      & MGFRec & 1.2996 & 1.5887 & 0.4232 \\
      & Agent4Rec & 0.8873 & 1.3494 & 0.4561 \\
      & Ours & \textbf{0.8272} & \textbf{1.3111} & \textbf{0.4623} \\
    \midrule
    \multirow{3}{*}{Baby}
      & MGFRec & 1.2973 & 1.5983 & 0.4399 \\
      & Agent4Rec & 0.8712 & 1.3489 & 0.4627 \\
      & Ours & \textbf{0.7999} & \textbf{1.2866} & \textbf{0.4685} \\
    \midrule
    \multirow{3}{*}{Book}
      & MGFRec & 1.5242 & 1.8187 & 0.4065 \\
      & Agent4Rec & 0.6914 & 1.0618 & 0.4446 \\
      & Ours & \textbf{0.5789} & \textbf{0.9387} & \textbf{0.4523} \\
    \bottomrule
  \end{tabular*}
\end{table}

\subsubsection{Feedback Simulation Quality}
To evaluate how well our simulator captures user preferences, we compare it with MGFRec~\cite{cai2026mgfrec} and Agent4Rec~\cite{zhang2024generative} using MAE and RMSE for rating prediction and BLEURT~\cite{sellam2020bleurt} for review generation. All simulators use the same backbone LLM and are evaluated on the validation and test sets. As shown in Table~\ref{tab:main_results}, our simulator achieves the lowest prediction errors and the highest BLEURT scores across the evaluated datasets. These results indicate better alignment with user ratings and reviews under the adopted metrics, supporting the effectiveness of the proposed feedback simulation approach in modeling preferences.

\subsubsection{Effect of Feedback Quality on Recommendation Performance}

\begin{figure}[t]
  \centering
  \captionsetup{skip=3pt}
  \includegraphics[width=\columnwidth]{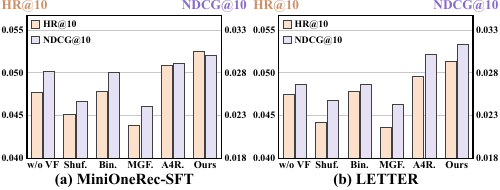}
  \caption{Comparison of different feedback settings on the Cell Phones dataset.}
  \label{fig:feedback_comp}
\end{figure}

To examine how different feedback strategies affect recommendation learning, we compare the following settings on MiniOneRec-SFT and LETTER. ``Bin.'' provides binary feedback derived from user interaction records, either ``I like it'' or ``I do not like it''. ``Shuf.'' replaces each user's feedback with that of a randomly selected user for the same candidate, disrupting personalized preference alignment. For ``MGF.'' (MGFRec) and ``A4R.'' (Agent4Rec), we replace the user simulator in our framework with their respective simulators to generate feedback, while ``Ours'' uses our proposed simulator.

Figure~\ref{fig:feedback_comp} compares the downstream recommendation performance of different feedback settings on both backbones. ``Bin.'' yields performance similar to ``w/o VF'', suggesting that binary feedback provides limited guidance for policy optimization. Notably, ``Shuf.'' and ``MGF.'' perform even worse than the setting without verbalized feedback, indicating that mismatched feedback can misguide candidate refinement and hinder policy learning. This further highlights that the effectiveness of verbalized feedback depends on its alignment with user preferences, rather than merely introducing textual guidance. The relative ordering of our simulator, ``A4R.'', and ``MGF.'' is consistent between the simulation evaluation and the downstream comparison. Better simulation scores are accompanied by stronger recommendation performance, suggesting that more accurate and informative verbalized feedback can provide more effective learning signals for recommendation models. 
% Together, these results support the use of personalized verbalized feedback to guide candidate refinement during RL training.

\subsection{Exploration during RL Training (RQ3)}

\begin{figure}[t]
  \centering
  \captionsetup{skip=3pt}
  \includegraphics[width=\columnwidth]{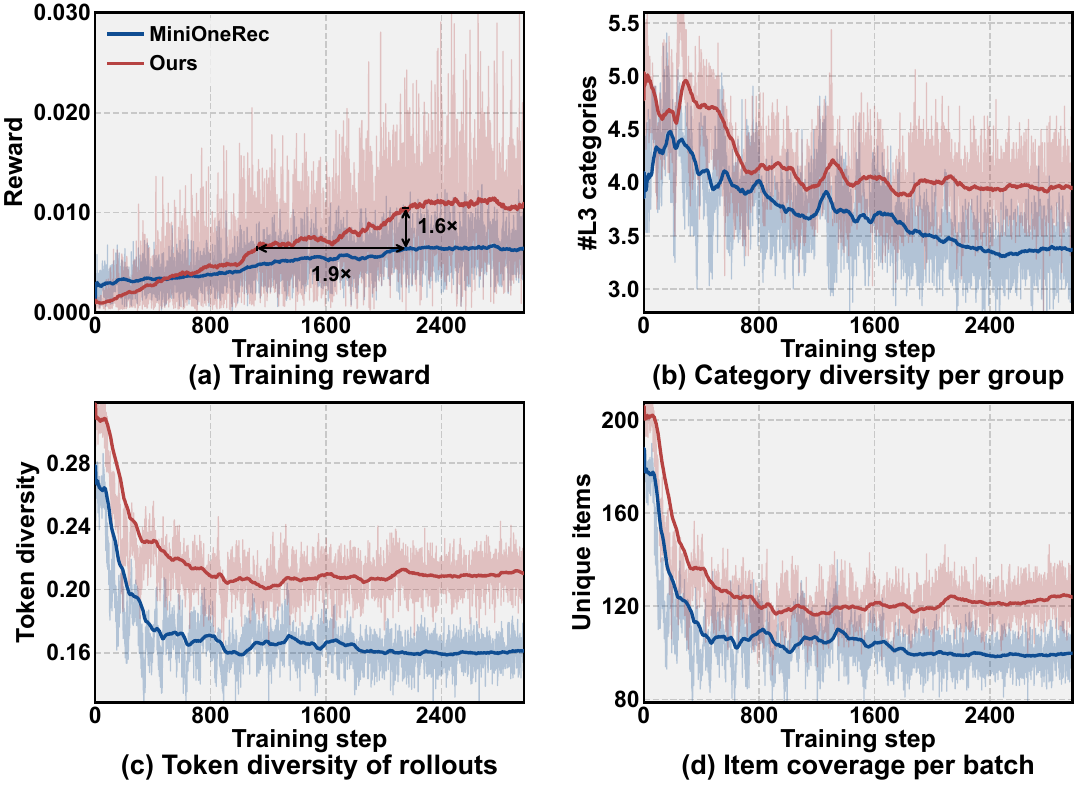}
  \caption{RL training dynamics on the Cell Phones dataset. We compare VerRec and MiniOneRec from the same MiniOneRec-SFT checkpoint.}
  \label{fig:reward_curve}
\end{figure}

\begin{figure*}[t]
  \centering
  \captionsetup{skip=3pt}
  \includegraphics[width=0.90\textwidth,keepaspectratio]{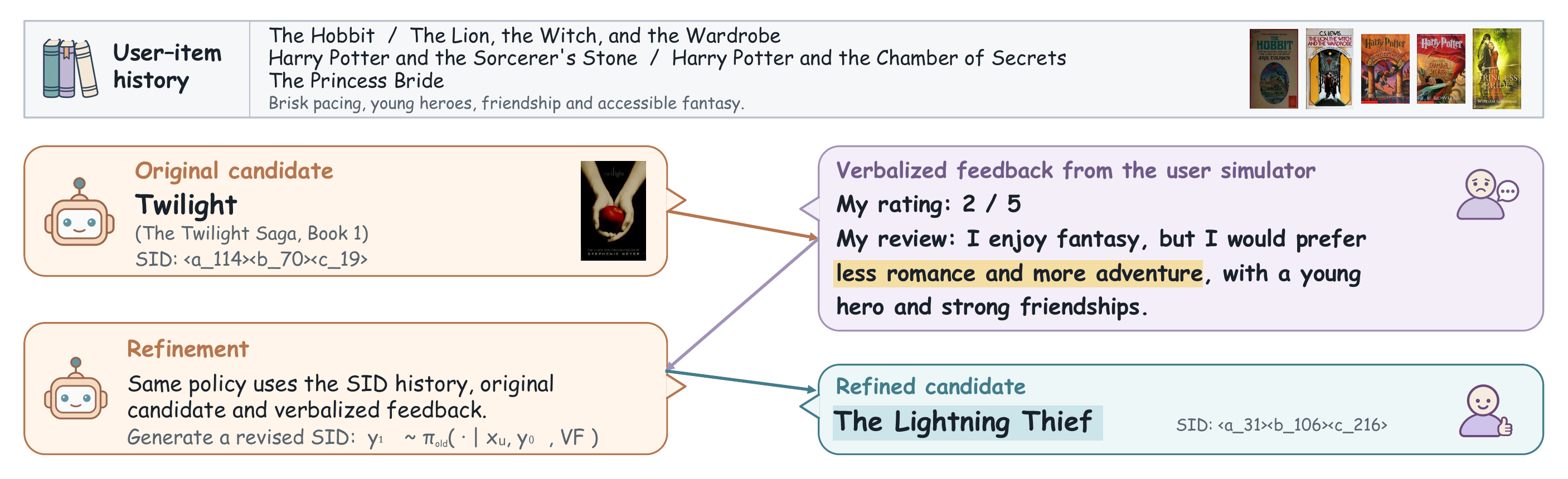}
  \caption{An example of how the recommendation policy revises its candidate through verbalized feedback.}
  \label{fig:book-refinement}
\end{figure*}
\subsubsection{Training Curves.}
To examine how verbalized feedback affects candidate exploration, we compare VerRec with MiniOneRec on Cell Phones, starting from the same MiniOneRec-SFT checkpoint.
Both methods generate and optimize over $2G$ candidates per request.
Figure~\ref{fig:reward_curve} presents training reward and three complementary exploration metrics: the number of distinct level-3 item categories per rollout group, SID token diversity, and the number of unique items per batch.

As shown in Figure~\ref{fig:reward_curve}(a), VerRec achieves greater reward improvements as training progresses. MiniOneRec requires approximately $1.9\times$ as many training steps to reach the reward level highlighted in the figure, while VerRec achieves approximately $1.6\times$ its reward at the indicated comparison step. VerRec achieves these gains at $1.47\times$ the total RL training time of the baseline.
These results suggest that feedback-guided refinement helps the policy identify candidates with higher rewards earlier in training, with moderate computational overhead.

Figures~\ref{fig:reward_curve}(b)--(d) further show that VerRec maintains greater exploration diversity across categories, SID tokens, and unique items. Although both methods exhibit an early decline in token diversity and unique items per batch, VerRec preserves a broader range of candidates as training proceeds. Its higher category diversity also indicates that this exploration extends across item categories. Together with the reward improvements, these observations suggest that personalized verbalized feedback provides useful directions for candidate refinement, helping the policy explore diverse alternatives that better satisfy the recommendation objective.

\subsubsection{Case Study.}
\label{sec:case_study}

To demonstrate the effectiveness of verbalized feedback, we present an illustrative example of refinement in the Book domain. As shown in Figure~\ref{fig:book-refinement}, the user's history includes \textit{The Hobbit}, \textit{The Lion, the Witch, and the Wardrobe}, two \textit{Harry Potter} books, and \textit{The Princess Bride}. These choices suggest a preference for fast-paced fantasy adventures with young heroes and strong friendships. The policy initially recommends \textit{Twilight}. The user simulator assigns it a rating of 2/5 and explains that the user prefers ``less romance and more adventure.'' Using the  history, original candidate, and verbalized feedback, the same policy generates a refined candidate corresponding to \textit{The Lightning Thief}. This recommendation better matches the preferences expressed in the feedback. The example illustrates how verbalized feedback explains a low rating and gives the policy an explicit direction for refinement.

\subsection{Impact of Key Components (RQ4)}
\subsubsection{Ablation Study.}
\label{sec:ablation}

\begin{figure}[t]
  \centering
  \captionsetup{skip=3pt}
  \includegraphics[width=\columnwidth]{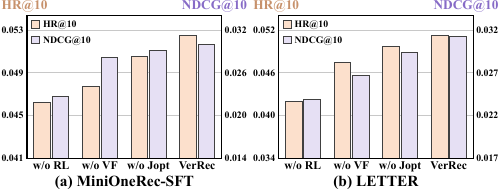}
  \caption{Ablation study on the Cell Phones dataset. “VF” and “Jopt” denote “Verbalized Feedback” and “Joint Optimization”, respectively.}
  \label{fig:ablation}
\end{figure}

To assess the contribution of each component, we conduct ablation studies of VerRec on Cell Phones using MiniOneRec-SFT and LETTER. The ``w/o RL'' variant denotes the original backbone before RL training. The ``w/o VF'' variant optimizes \(2G\)  candidates per input on the corresponding backbone without verbalized feedback. The ``w/o Jopt'' variant excludes original candidates from optimization and trains only on refined candidates.

As shown in Figure~\ref{fig:ablation}, VerRec achieves the best performance on both metrics across the two backbones. Removing verbalized feedback reduces performance, suggesting that explicit guidance about preference mismatches and revision directions benefits learning beyond scalar rewards alone. Removing joint optimization also reduces performance, although training on refined candidates still improves over the original backbones. This indicates the benefit of jointly learning from original and refined candidates. Together, these results support the contributions of verbalized feedback and joint rollout optimization to recommendation performance.

\subsubsection{Hyperparameter Analysis.}
\label{sec:hyperparameter}

We examine how the number of candidates selected for refinement and the original candidate group size $G$ affect recommendation performance on the Cell Phones dataset. 

\begin{figure}[t]
  \centering
  \captionsetup{skip=3pt}
  \includegraphics[width=\columnwidth]{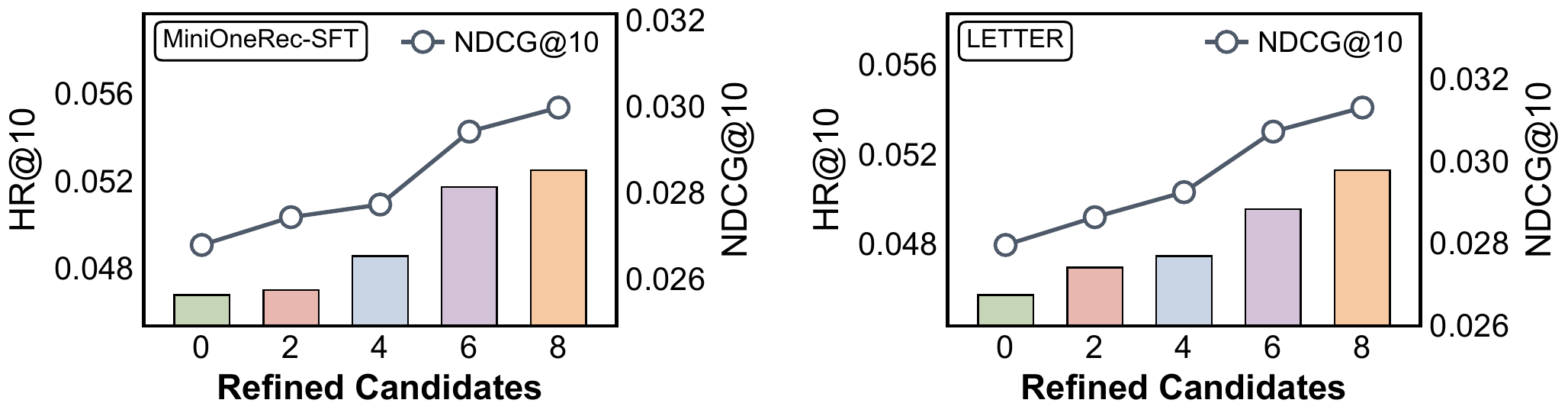}
  \caption{Effect of the number of candidates selected for the refinement on the Cell Phones dataset.}
  \label{fig:refine}
\end{figure}

\textbf{Number of candidates selected for refinement.}
We vary the number of original candidates selected for refinement among $\{0,2,4,6,8\}$. As shown in Figure~\ref{fig:refine}, both evaluation metrics improve on MiniOneRec-SFT and LETTER as more candidates are refined. This trend suggests that applying verbalized feedback to more candidates creates more opportunities to find rewarding recommendations, providing useful learning signals for optimization.

\begin{figure}[t]
  \centering
  \captionsetup{skip=3pt}
  \includegraphics[width=\columnwidth]{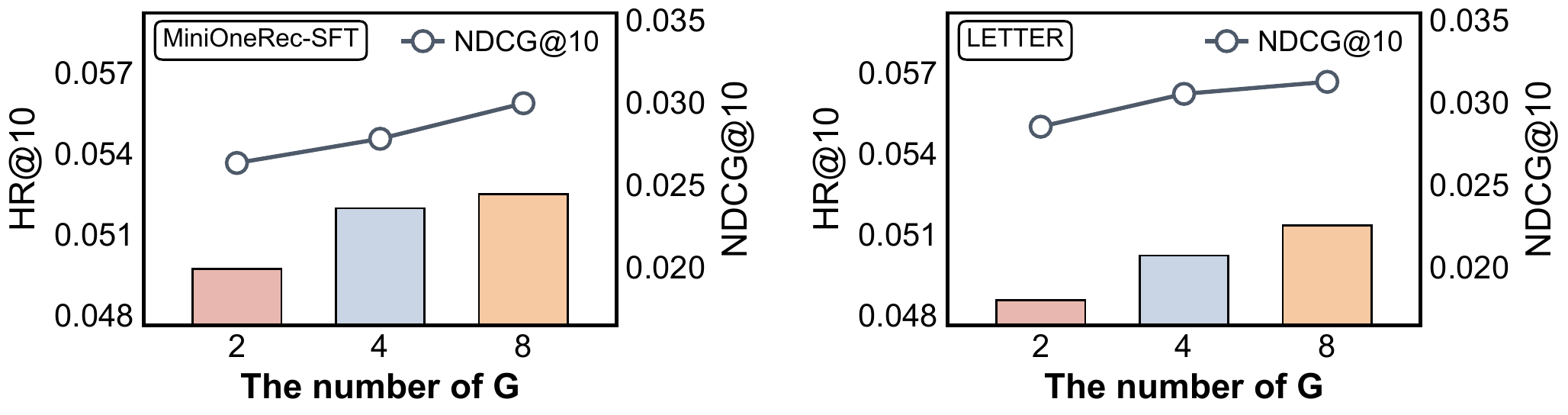}
  \caption{Effect of the number of generated candidates G on the Cell Phones dataset.}
  \label{fig:beam}
\end{figure}

\textbf{Rollout group size.}
We vary $G$ among $\{2,4,8\}$ and generate one refined candidate for each original candidate, yielding an expanded group of size $2G$. As shown in Figure~\ref{fig:beam}, increasing $G$ improves both metrics on MiniOneRec-SFT and LETTER, with $G=8$ achieving the best performance among the tested settings. A larger group provides more candidates for exploration and more opportunities to learn from reward differences. 

\section{Conclusion and Future Work}
\label{sec:conclusion}

In this work, we propose VerRec, a novel reinforcement learning framework with verbalized feedback for generative recommendation. VerRec augments scalar rewards with personalized feedback generated by an LLM-based user simulator.  Experiments on three Amazon review datasets with two backbones demonstrate that VerRec improves recommendation performance. Further analyses highlight the complementary roles of personalized verbalized feedback and joint optimization in guiding exploration toward more diverse candidates with higher rewards.

In future work, we will develop more personalized and accurate user simulators to improve feedback quality. Moreover, incorporating real-world users' verbalized feedback into this framework is a promising direction for guiding candidate refinement and aligning the policy with users’ evolving preferences. Finally, we will also investigate more efficient feedback generation (e.g., asynchronous generation) to reduce RL training overhead.

%% Ethical Considerations (encouraged by WWW 2027 for ethical use of data / informed consent)
% \input{latex/7_ethical_considerations}

%% --- Bibliography --------------------------------------------------------
\bibliographystyle{ACM-Reference-Format}
\bibliography{ref}

@article{rajput2023recommender,
  title={Recommender systems with generative retrieval},
  author={Rajput, Shashank and Mehta, Nikhil and Singh, Anima and Hulikal Keshavan, Raghunandan and Vu, Trung and Heldt, Lukasz and Hong, Lichan and Tay, Yi and Tran, Vinh and Samost, Jonah and others},
  journal={Advances in Neural Information Processing Systems},
  volume={36},
  pages={10299--10315},
  year={2023}
}

@article{kong2025minionerec,
  title={Minionerec: An open-source framework for scaling generative recommendation},
  author={Kong, Xiaoyu and Sheng, Leheng and Tan, Junfei and Chen, Yuxin and Wu, Jiancan and Zhang, An and Wang, Xiang and He, Xiangnan},
  journal={arXiv preprint arXiv:2510.24431},
  year={2025}
}

@article{tan2025rere,
  title={Reinforced preference optimization for recommendation},
  author={Tan, Junfei and Chen, Yuxin and Zhang, An and Jiang, Junguang and Liu, Bin and Xu, Ziru and Zhu, Han and Xu, Jian and Zheng, Bo and Wang, Xiang},
  journal={arXiv preprint arXiv:2510.12211},
  year={2025}
}

@article{shao2024deepseekmath,
  title={Deepseekmath: Pushing the limits of mathematical reasoning in open language models},
  author={Shao, Zhihong and Wang, Peiyi and Zhu, Qihao and Xu, Runxin and Song, Junxiao and Bi, Xiao and Zhang, Haowei and Zhang, Mingchuan and Li, YK and Wu, Yang and others},
  journal={arXiv preprint arXiv:2402.03300},
  year={2024}
}

@inproceedings{kang2018self,
  title={Self-attentive sequential recommendation},
  author={Kang, Wang-Cheng and McAuley, Julian},
  booktitle={2018 IEEE international conference on data mining (ICDM)},
  pages={197--206},
  year={2018},
  organization={IEEE}
}

@article{hidasi2015session,
  title={Session-based recommendations with recurrent neural networks},
  author={Hidasi, Bal{\'a}zs and Karatzoglou, Alexandros and Baltrunas, Linas and Tikk, Domonkos},
  journal={arXiv preprint arXiv:1511.06939},
  year={2015}
}

@inproceedings{tang2018personalized,
  title={Personalized top-n sequential recommendation via convolutional sequence embedding},
  author={Tang, Jiaxi and Wang, Ke},
  booktitle={Proceedings of the eleventh ACM international conference on web search and data mining},
  pages={565--573},
  year={2018}
}

@article{bao2025bi,
  title={A bi-step grounding paradigm for large language models in recommendation systems},
  author={Bao, Keqin and Zhang, Jizhi and Wang, Wenjie and Zhang, Yang and Yang, Zhengyi and Luo, Yanchen and Chen, Chong and Feng, Fuli and Tian, Qi},
  journal={ACM Transactions on Recommender Systems},
  volume={3},
  number={4},
  pages={1--27},
  year={2025},
  publisher={ACM New York, NY}
}

@article{bao2024decoding,
  title={Decoding matters: Addressing amplification bias and homogeneity issue for llm-based recommendation},
  author={Bao, Keqin and Zhang, Jizhi and Zhang, Yang and Huo, Xinyue and Chen, Chong and Feng, Fuli},
  journal={arXiv preprint arXiv:2406.14900},
  year={2024}
}

@article{chen2024softmax,
  title={On softmax direct preference optimization for recommendation},
  author={Chen, Yuxin and Tan, Junfei and Zhang, An and Yang, Zhengyi and Sheng, Leheng and Zhang, Enzhi and Wang, Xiang and Chua, Tat-Seng},
  journal={Advances in Neural Information Processing Systems},
  volume={37},
  pages={27463--27489},
  year={2024}
}

@inproceedings{wang2024learnable,
  title={Learnable item tokenization for generative recommendation},
  author={Wang, Wenjie and Bao, Honghui and Lin, Xinyu and Zhang, Jizhi and Li, Yongqi and Feng, Fuli and Ng, See-Kiong and Chua, Tat-Seng},
  booktitle={Proceedings of the 33rd ACM International Conference on Information and Knowledge Management},
  pages={2400--2409},
  year={2024}
}

@article{you2026r,
  title={{R$^{2}$ec}: Towards Large Recommender Models with Reasoning},
  author={You, Runyang and Li, Yongqi and Lin, Xinyu and Zhang, Xin and Wang, Wenjie and Li, Wenjie and Nie, Liqiang},
  journal={Advances in Neural Information Processing Systems},
  volume={38},
  pages={62376--62405},
  year={2026}
}

@inproceedings{zhang2026reinforced,
  title={Reinforced latent reasoning for llm-based recommendation},
  author={Zhang, Yang and Xu, Wenxin and Zhao, Xiaoyan and Wang, Wenjie and Feng, Fuli and He, Xiangnan and Chua, Tat-Seng},
  booktitle={International Conference on Learning Representations},
  volume={2026},
  pages={128449--128470},
  year={2026}
}

@inproceedings{zheng2024adapting,
  title={Adapting large language models by integrating collaborative semantics for recommendation},
  author={Zheng, Bowen and Hou, Yupeng and Lu, Hongyu and Chen, Yu and Zhao, Wayne Xin and Chen, Ming and Wen, Ji-Rong},
  booktitle={2024 IEEE 40th International Conference on Data Engineering (ICDE)},
  pages={1435--1448},
  year={2024},
  organization={IEEE}
}

@article{zhou2025onerec,
  title={Onerec technical report},
  author={Zhou, Guorui and Deng, Jiaxin and Zhang, Jinghao and Cai, Kuo and Ren, Lejian and Luo, Qiang and Wang, Qianqian and Hu, Qigen and Huang, Rui and Wang, Shiyao and others},
  journal={arXiv preprint arXiv:2506.13695},
  year={2025}
}

@inproceedings{cai2026mgfrec,
  title={Mgfrec: Towards reinforced reasoning recommendation with multiple groundings and feedback},
  author={Cai, Shihao and Gao, Chongming and Liu, Haoyan and Shi, Wentao and Sun, Jianshan and Tang, Ruiming and Feng, Fuli},
  booktitle={Proceedings of the 32nd ACM SIGKDD Conference on Knowledge Discovery and Data Mining V. 1},
  pages={49--58},
  year={2026}
}

@article{feng2026fine,
  title={Fine-grained Semantics Integration for Large Language Model-based Recommendation},
  author={Feng, Jiawei and Kong, Xiaoyu and Sheng, Leheng and Wu, Bin and Yi, Chao and Yang, Feifang and Sheng, Xiang-Rong and Zhu, Han and Wang, Xiang and Wu, Jiancan and others},
  journal={arXiv preprint arXiv:2602.22632},
  year={2026}
}

@article{liu2025onerec,
  title={Onerec-think: In-text reasoning for generative recommendation},
  author={Liu, Zhanyu and Wang, Shiyao and Wang, Xingmei and Zhang, Rongzhou and Deng, Jiaxin and Bao, Honghui and Zhang, Jinghao and Li, Wuchao and Zheng, Pengfei and Wu, Xiangyu and others},
  journal={arXiv preprint arXiv:2510.11639},
  year={2025}
}

@article{zhou2025openonerec,
  title={Openonerec technical report},
  author={Zhou, Guorui and Bao, Honghui and Huang, Jiaming and Deng, Jiaxin and Zhang, Jinghao and She, Junda and Cai, Kuo and Ren, Lejian and Ren, Lu and Luo, Qiang and others},
  journal={arXiv preprint arXiv:2512.24762},
  year={2025}
}

@article{ding2026towards,
  title={Towards Sample-Efficient and Stable Reinforcement Learning for LLM-based Recommendation},
  author={Ding, Hongxun and Bao, Keqin and Zhang, Jizhi and Fang, Yi and Xu, Wenxin and Feng, Fuli and He, Xiangnan},
  journal={arXiv preprint arXiv:2602.00632},
  year={2026}
}

@inproceedings{wang2026gflowgr,
  title={Gflowgr: Fine-tuning generative recommendation frameworks with generative flow networks},
  author={Wang, Yejing and Zhou, Shengyu and Lu, Jinyu and Liu, Qidong and Li, Xinhang and Zhang, Wenlin and Li, Feng and Wang, Pengjie and Yu, Chuan and Xu, Jian and others},
  booktitle={Proceedings of the 49th International ACM SIGIR Conference on Research and Development in Information Retrieval},
  pages={1925--1936},
  year={2026}
}

@article{fang2026reason4rec,
  title={Reason4Rec: Deliberative User Preference Alignment of Large Language Models for Recommendation},
  author={Fang, Yi and Wang, Wenjie and Zhang, Yang and Zhu, Fengbin and Wang, Qifan and Feng, Fuli and He, Xiangnan},
  journal={IEEE Transactions on Knowledge and Data Engineering},
  year={2026},
  publisher={IEEE}
}

@inproceedings{zhang2024generative,
  title={On generative agents in recommendation},
  author={Zhang, An and Chen, Yuxin and Sheng, Leheng and Wang, Xiang and Chua, Tat-Seng},
  booktitle={Proceedings of the 47th international ACM SIGIR conference on research and development in Information Retrieval},
  pages={1807--1817},
  year={2024}
}

@article{yang2025qwen3,
  title={Qwen3 technical report},
  author={Yang, An and Li, Anfeng and Yang, Baosong and Zhang, Beichen and Hui, Binyuan and Zheng, Bo and Yu, Bowen and Gao, Chang and Huang, Chengen and Lv, Chenxu and others},
  journal={arXiv preprint arXiv:2505.09388},
  year={2025}
}

@misc{qwen2.5,
    title = {Qwen2.5: A Party of Foundation Models},
    url = {https://qwenlm.github.io/blog/qwen2.5/},
    author = {Qwen Team},
    month = {September},
    year = {2024}
}

@inproceedings{sellam2020bleurt,
  title={BLEURT: Learning robust metrics for text generation},
  author={Sellam, Thibault and Das, Dipanjan and Parikh, Ankur},
  booktitle={Proceedings of the 58th annual meeting of the association for computational linguistics},
  pages={7881--7892},
  year={2020}
}

@article{team2026onereason,
  title={OneReason Technical Report},
  author={Team, OneRec and Yang, Biao and Ding, Boyang and Chu, Chenglong and Zang, Dunju and Pan, Fei and Li, Han and Jiang, Hao and Bao, Honghui and Wang, Huanjie and others},
  journal={arXiv preprint arXiv:2606.06260},
  year={2026}
}

@article{chen2026beyond,
  title={Beyond Action Imitation: Learning a Decision-Aware User Simulator for Online Advertising},
  author={Chen, Zipeng and Zheng, Jiaer and Xu, Xiangyang and Lin, Xinyu and Wang, Zhaobin and Liu, Zhaohui and Xiang, Qianjin and Zhao, Xiaoyu and Yu, Zhuozhen and Wang, Guangshuo and others},
  journal={arXiv preprint arXiv:2607.26893},
  year={2026}
}

@article{zhang2025critique,
  title={Critique-grpo: Advancing llm reasoning with natural language and numerical feedback},
  author={Zhang, Xiaoying and Zhang, Yipeng and Sun, Hao and Feng, Kaituo and Lu, Chaochao and Yang, Chao and Meng, Helen},
  journal={arXiv preprint arXiv:2506.03106},
  year={2025}
}

@article{huang2026bootstrapping,
  title={Bootstrapping exploration with group-level natural language feedback in reinforcement learning},
  author={Huang, Lei and Cheng, Xiang and Zhao, Chenxiao and Shen, Guobin and Yang, Junjie and Feng, Xiaocheng and Gu, Yuxuan and Yu, Xing and Qin, Bing},
  journal={arXiv preprint arXiv:2603.04597},
  year={2026}
}

@inproceedings{he2026reasoning,
  title={Reasoning over semantic ids enhances generative recommendation},
  author={He, Yingzhi and Sun, Yan and Tan, Junfei and Chen, Yuxin and Kong, Xiaoyu and Shen, Chunxu and Wang, Xiang and Zhang, An and Chua, Tat-Seng},
  booktitle={Proceedings of the 32nd ACM SIGKDD Conference on Knowledge Discovery and Data Mining V. 2},
  pages={1638--1649},
  year={2026}
}

@inproceedings{geng2022recommendation,
  title={Recommendation as language processing (rlp): A unified pretrain, personalized prompt \& predict paradigm (p5)},
  author={Geng, Shijie and Liu, Shuchang and Fu, Zuohui and Ge, Yingqiang and Zhang, Yongfeng},
  booktitle={Proceedings of the 16th ACM conference on recommender systems},
  pages={299--315},
  year={2022}
}

@article{zhang2026recommendation,
  title={Recommendation as instruction following: A large language model empowered recommendation approach},
  author={Zhang, Junjie and Xie, Ruobing and Hou, Yupeng and Zhao, Wayne Xin and Lin, Leyu and Wen, Ji-Rong},
  journal={ACM Transactions on Information Systems},
  volume={43},
  number={5},
  pages={1--37},
  year={2026},
  publisher={ACM New York, NY}
}

@inproceedings{ye2026align3gr,
  title={Align$^3$GR: Unified Multi-Level Alignment for LLM-based Generative Recommendation},
  author={Ye, Wencai and Sun, Mingjie and Chen, Shuhang and Wu, Wenjin and Jiang, Peng},
  booktitle={Proceedings of the AAAI Conference on Artificial Intelligence},
  volume={40},
  number={19},
  pages={16154--16162},
  year={2026}
}

@article{madaan2023self,
  title={Self-refine: Iterative refinement with self-feedback},
  author={Madaan, Aman and Tandon, Niket and Gupta, Prakhar and Hallinan, Skyler and Gao, Luyu and Wiegreffe, Sarah and Alon, Uri and Dziri, Nouha and Prabhumoye, Shrimai and Yang, Yiming and others},
  journal={Advances in neural information processing systems},
  volume={36},
  pages={46534--46594},
  year={2023}
}

@article{shinn2023reflexion,
  title={Reflexion: Language agents with verbal reinforcement learning},
  author={Shinn, Noah and Cassano, Federico and Gopinath, Ashwin and Narasimhan, Karthik and Yao, Shunyu},
  journal={Advances in neural information processing systems},
  volume={36},
  pages={8634--8652},
  year={2023}
}

@inproceedings{gou2024critic,
  title={Critic: Large language models can self-correct with tool-interactive critiquing},
  author={Gou, Zhibin and Shao, Zhihong and Gong, Yeyun and Yang, Yujiu and Duan, Nan and Chen, Weizhu and others},
  booktitle={International Conference on Learning Representations},
  volume={2024},
  pages={57734--57811},
  year={2024}
}

@inproceedings{akyurek2023rl4f,
  title={Rl4f: Generating natural language feedback with reinforcement learning for repairing model outputs},
  author={Aky{\"u}rek, Afra Feyza and Aky{\"u}rek, Ekin and Kalyan, Ashwin and Clark, Peter and Wijaya, Derry Tanti and Tandon, Niket},
  booktitle={Proceedings of the 61st Annual Meeting of the Association for Computational Linguistics (Volume 1: Long Papers)},
  pages={7716--7733},
  year={2023}
}

@article{stephan2024rlvf,
  title={Rlvf: Learning from verbal feedback without overgeneralization},
  author={Stephan, Moritz and Khazatsky, Alexander and Mitchell, Eric and Chen, Annie S and Hsu, Sheryl and Sharma, Archit and Finn, Chelsea},
  journal={arXiv preprint arXiv:2402.10893},
  year={2024}
}

@inproceedings{chen2024teaching,
  title={Teaching large language models to self-debug},
  author={Chen, Xinyun and Lin, Maxwell and Sch{\"a}rli, Nathanael and Zhou, Denny},
  booktitle={International Conference on Learning Representations},
  volume={2024},
  pages={8746--8825},
  year={2024}
}

@inproceedings{paul2024refiner,
  title={Refiner: Reasoning feedback on intermediate representations},
  author={Paul, Debjit and Ismayilzada, Mete and Peyrard, Maxime and Borges, Beatriz and Bosselut, Antoine and West, Robert and Faltings, Boi},
  booktitle={Proceedings of the 18th Conference of the European Chapter of the Association for Computational Linguistics (Volume 1: Long Papers)},
  pages={1100--1126},
  year={2024}
}

@article{ie2019recsim,
  title={Recsim: A configurable simulation platform for recommender systems},
  author={Ie, Eugene and Hsu, Chih-wei and Mladenov, Martin and Jain, Vihan and Narvekar, Sanmit and Wang, Jing and Wu, Rui and Boutilier, Craig},
  journal={arXiv preprint arXiv:1909.04847},
  year={2019}
}

@article{zhao2023kuaisim,
  title={KuaiSim: A comprehensive simulator for recommender systems},
  author={Zhao, Kesen and Liu, Shuchang and Cai, Qingpeng and Zhao, Xiangyu and Liu, Ziru and Zheng, Dong and Jiang, Peng and Gai, Kun},
  journal={Advances in Neural Information Processing Systems},
  volume={36},
  pages={44880--44897},
  year={2023}
}

@inproceedings{shi2019virtual,
  title={Virtual-taobao: Virtualizing real-world online retail environment for reinforcement learning},
  author={Shi, Jing-Cheng and Yu, Yang and Da, Qing and Chen, Shi-Yong and Zeng, An-Xiang},
  booktitle={Proceedings of the AAAI Conference on Artificial Intelligence},
  volume={33},
  number={01},
  pages={4902--4909},
  year={2019}
}

@article{wang2025user,
  title={User behavior simulation with large language model-based agents},
  author={Wang, Lei and Zhang, Jingsen and Yang, Hao and Chen, Zhi-Yuan and Tang, Jiakai and Zhang, Zeyu and Chen, Xu and Lin, Yankai and Sun, Hao and Song, Ruihua and others},
  journal={ACM Transactions on Information Systems},
  volume={43},
  number={2},
  pages={1--37},
  year={2025},
  publisher={ACM New York, NY}
}

@inproceedings{chen2025recusersim,
  title={Recusersim: A realistic and diverse user simulator for evaluating conversational recommender systems},
  author={Chen, Luyu and Dai, Quanyu and Zhang, Zeyu and Feng, Xueyang and Zhang, Mingyu and Tang, Pengcheng and Chen, Xu and Zhu, Yue and Dong, Zhenhua},
  booktitle={Companion Proceedings of the ACM on Web Conference 2025},
  pages={133--142},
  year={2025}
}

@inproceedings{zhang2026exploring,
  title={Exploring Recommender System Evaluation: A Multi-Modal LLM Agent Framework for A/B Testing},
  author={Zhang, Wenlin and Li, Xiangyang and Ge, Qiyuan and Dong, Kuicai and Jia, Pengyue and Li, Xiaopeng and Zhang, Zijian and Wang, Maolin and Wang, Yichao and Guo, Huifeng and others},
  booktitle={Proceedings of the 32nd ACM SIGKDD Conference on Knowledge Discovery and Data Mining V. 1},
  pages={2878--2889},
  year={2026}
}

@inproceedings{zhang2026shop,
  title={Shop-r1: Rewarding llms to simulate human behavior in online shopping via reinforcement learning},
  author={Zhang, Yimeng and Wang, Tian and Gesi, Jiri and Wang, Ziyi and Lu, Yuxuan and Lin, Jiacheng and Zhan, Simon and Gao, Vianne and Jiao, Ruochen and Liu, Junze and others},
  booktitle={International Conference on Learning Representations},
  volume={2026},
  pages={8406--8423},
  year={2026}
}

@inproceedings{liu2026recoworld,
  title={Recoworld: Building simulated environments for agentic recommender systems},
  author={Liu, Fei and Lin, Xinyu and Yu, Hanchao and Wu, Mingyuan and Wang, Jianyu and Zhang, Qiang and Zhao, Zhuokai and Xia, Yinglong and Zhang, Yao and Li, Weiwei and others},
  booktitle={Companion Proceedings of the ACM Web Conference 2026},
  pages={650--659},
  year={2026}
}

@inproceedings{wei2026mirroring,
  title={Mirroring Users: Towards Building Preference-aligned User Simulator with User Feedback in Recommendation},
  author={Wei, Tianjun and Guo, Huizhong and Du, Yingpeng and Sun, Zhu and Chen, Huang and Wang, Dongxia and Zhang, Jie},
  booktitle={Proceedings of the 64th Annual Meeting of the Association for Computational Linguistics (Volume 1: Long Papers)},
  pages={6046--6071},
  year={2026}
}

@inproceedings{yoon2024evaluating,
  title={Evaluating large language models as generative user simulators for conversational recommendation},
  author={Yoon, Se-eun and He, Zhankui and Echterhoff, Jessica and McAuley, Julian},
  booktitle={Proceedings of the 2024 Conference of the North American Chapter of the Association for Computational Linguistics: Human Language Technologies (Volume 1: Long Papers)},
  pages={1490--1504},
  year={2024}
}

@inproceedings{zhu2024reliable,
  title={How reliable is your simulator? analysis on the limitations of current llm-based user simulators for conversational recommendation},
  author={Zhu, Lixi and Huang, Xiaowen and Sang, Jitao},
  booktitle={Companion Proceedings of the ACM Web Conference 2024},
  pages={1726--1732},
  year={2024}
}

@inproceedings{lin2026bringing,
  title={Bringing reasoning to generative recommendation through the lens of cascaded ranking},
  author={Lin, Xinyu and Liu, Pengyuan and Wang, Wenjie and Hu, Yicheng and Xu, Chen and Feng, Fuli and Wang, Qifan and Chua, Tat-Seng},
  booktitle={Proceedings of the ACM Web Conference 2026},
  pages={8939--8949},
  year={2026}
}

@article{yan2026learning,
  title={Learning to reason under off-policy guidance},
  author={Yan, Jianhao and Li, Yafu and Hu, Zican and Wang, Zhi and Cui, Ganqu and Qu, Xiaoye and Cheng, Yu and Zhang, Yue},
  journal={Advances in Neural Information Processing Systems},
  volume={38},
  pages={117157--117186},
  year={2026}
}

@inproceedings{fan2026uncertainty,
  title={Uncertainty-aware generative recommendation},
  author={Fan, Chenxiao and Gao, Chongming and Gong, Yaxin and Liu, Haoyan and Feng, Fuli and He, Xiangnan},
  booktitle={Proceedings of the 32nd ACM SIGKDD Conference on Knowledge Discovery and Data Mining V. 2},
  pages={1015--1026},
  year={2026}
}

%% --- Appendix ------------------------------------------------------------
% \appendix
% \input{latex/appendix_contrastive}
% % \input{latex/appendix}

\end{document}